\documentclass{ws-ijgmmp}
\usepackage[colorlinks=true]{hyperref}
\usepackage[numbers,sort&compress]{natbib}
\usepackage{float}
\usepackage[subrefformat=parens, caption=false]{subfig}
\begin{document}

\title{Nonperturbative thermodynamic extrinsic curvature  of the anyon gas
}

\author{{Mahnaz Tavakoli Kachi \footnote{m.tavakoli1386@ph.iut.ac.ir}}, {Behrouz Mirza \footnote{b.mirza@iut.ac.ir}} and Fatemeh Sadat Hashemi \footnote{hashemifatemeh038@gmail.com} \\}

\address{Department of Physics,\\ Isfahan University of Technology,\\ Isfahan 84156-83111, Iran}

\maketitle


\begin{abstract}
Thermodynamic extrinsic curvature is a new  mathematical tool in thermodynamic geometry. By using the thermodynamic extrinsic curvature,  one may obtain a more complete geometric representation of the critical phenomena and thermodynamics. We introduce  nonperturbative thermodynamic extrinsic curvature of an ideal two dimensional gas of anyons.  Using extrinsic curvature, we find new fixed  points   in  nonperturbative thermodynamics   of the anyon gas that particles behave as semions.   Here, we investigate the critical behavior of  thermodynamic extrinsic curvature of two-dimensional Kagome Ising model near  the critical point  $ \beta_{c}   =({{k_{B} T_{c}}})^{-1}$ in a constant magnetic field and show that it behaves as $ \left| {\beta- \beta_{c} } \right|^{\alpha}   $ with  $ \alpha=0 $, where $  \alpha $ denotes the critical exponent of the specific heat.  Then,  we consider the three dimensional  spherical model and show that  the scaling behavior is $ \left| {\beta- \beta_{c} } \right|^{\alpha}   $ , where $ \alpha =-1 $. Finally, using a   general  argument, we show that extrinsic curvature $ K $  have two different scaling behaviors for positive and negative $ \alpha $. For  $\alpha> 0$, our results indicate that   $ K \sim \left|{\beta- \beta_{c} } \right|^{{\frac{1}{2}} (\alpha-2)} $. However, for  $ \alpha <0$, we found  a different scaling behavior, where  $ K\sim  \left| {\beta- \beta_{c} } \right|^{\alpha} $.
\end{abstract}


\section{Introduction}
Weinhold introduced a  geometrical metric based on the Hessian matrix  of the internal energy \cite{Weinhold}. However, he
did not investigate the distance between the thermodynamic states.   Ruppeiner introduced a metric and suggested a correspondence between  the singularities that appears in thermodynamic geometry and phase transitions \cite{Ruppeiner,geometry}. A proof of the    correspondence  was proposed  in \cite{Mirza, Fazel}. It was suggested by Ruppeiner that  the  sign of the thermodynamic scalar curvature may determine  the character of the microscopic interaction, whether it is repulsive ($ R<0 $) or attractive ($R>0$).  The thermodynamic  scalar curvature $ R $ of the ideal gas is equal to zero ($ R=0 $) \cite{Ruppeiner}. The scalar curvature can be used as a new tool to measure the amount of interaction among microstates.

So far, the thermodynamic geometry of various sytems has been obtained.  For a review of those models, see \cite{Rupplist1}. Janyszek and Mrugala have obtained the  thermodynamic scalar curvature of ideal quantum gases \cite{Janyszek}. The results show that $ R $  is negative for a gas of Fermions  and is positive for a Bose gas.    Thermodynamic geometry of an anyon gas which obeys fractional statistics has been investigated  via the thermodynamic scalar  curvature $ R $   and  some notable aspects  of anyon gas has been reported \cite{anyon1,Mohammadzadeh}. The scalar curvature of anyons has both   signs,  whose   change from  positive  (Bose like behavior) to   negative values (Fermi like gas).   

Near the critical point, both correlation length $\xi$ and the thermodynamic curvature $R$ diverge.  It was  suggested and confirmed in different studies  that near the critical point the thermodynamic scalar curvature can be written as $ R\sim\kappa \,\xi^{{d}}$, where  ${d}$ is  dimension of the system, $\xi$  is the correlation length  and $ \kappa $ is a constant \cite{kagome,Ruppeiner,geometry,s1,s2,s3,Nakamura}.   Considering  the scaling relation  $ \nu d=2-\alpha $  where $\alpha$ and $\nu$  are critical exponents,   the thermodynamic curvature behaves as  $ 	R \sim t^{\alpha-2} $ where $ t $ denotes the difference in  the temperature from its critical value,   $ t= \beta- \beta_{c} $.  Therefore, using the scaling behavior of $ R $, the critical exponent of the  thermodynamic  system can be investigated, e.g.  the scaling behavior of the  thermodynamic scalar curvature for three dimensional Van-der  Walls model and four dimensional spherical model  is  $ 	R \sim t^{\alpha-2} $ ($ \alpha=0 $) \cite{spheric}. While the critical behavior   of  thermodynamic scalar curvature  of two dimensional Kagome Ising model  is $ R  \sim t^{\alpha-1} $, where $ \alpha=0 $ \cite{kagome}.  Therefore, we expect that in some cases the scaling behavior of $ R $ depends on the dimension of  thermodynamic system.

Thermodynamic geometry of black holes may give us useful information  about their microstates interactions \cite{Geometrothermodynamics}.  It was suggested that microstes in the Kerr-Newman black holes might behave as fermions and represent repulsive  interactions \cite{blackhole2,kerr}. Also,  it was found that the critical exponents of the Van der Waals fluid are the same as that of the charged AdS black holes \cite{Universal1,hairyblackholes,Universal}.

Among these,  an interesting subject is to investigate thermodynamic extrinsic curvature in statistical systems. The inspiration of this work is  the earlier study of   black holes  thermodynamic geometry  by using  the extrinsic curvature \cite{Mansoori}. The results indicate the correspondence   between the singularities  of the  thermodynamic extrinsic curvature and the phase transition points  that are corresponded to  the   heat capacity.  The thermodynamic extrinsic curvature could be an important geometrical quantity in the study of critical phenomena. In this work we start to investigate the following issues. Whether the extrinsic curvature on an embedded hypersurface in the thermodynamic manifold can provide useful  information about  critical phenomenon. Does the scaling behavior of the thermodynamic extrinsic curvature  depends on the dimension? What type of universal behaviors and information may be obtained study of the thermodynamic extrinsic curvature.  To explore usefulness of thermodynamic extrinsic curvature, we study some statistical systems in the following sections.
In this paper, we  study  the critical behavior of the gas of anyons and two critical spin systems by using thermodynamic extrinsic curvature. We derive the thermodynamic extrinsic curvature and compare its properties with the well known thermodynamic Ricci scalar.   

This paper is outlined as follows. In section \ref{sec:geo}, we   review some of the basic relations  in the thermodynamic geometry. In section  \ref{sec:anyon}, we first briefly present the thermodynamic properties of an anyon gas which obeys fractional statistics and obtain it's thermodynamic  extrinsic curvature.  In section  \ref{sec:critic}, we  consider  thermodynamic geometry of two dimensional Kagome Ising model and find  the critical exponent   using the extrinsic curvature. We also explore  thermodynamic geometry of three dimensional   spherical model. The standard scaling behavior of the extrinsic curvature  is given in  \ref{sec:standard}. A summary of the results can be found in section   \ref{sec:conclud}.  Details on relations  of the Kagome Ising model can be reached in Appendix.

\section{ Thermodynamic geometry}\label{sec:geo}
Weinhold and Ruppeiner introduced a geometrical formulation of thermodynamic \cite{Weinhold,Ruppeiner}.   Weinhold suggested the energy representation by using  the  internal energy  and its second  derivative.  Ruppeiner introduced the metric by the second derivative of entropy. It was suggested that both formulations  are  equivalent  and the metrics are related by a conformal transformation \cite{Salamon}.  Using  Legendre transformations one may generate  other forms of thermodynamic  metrics. A new form of thermodynamic  metric was  recently proposed in \cite{Mirza}.  In 1990, Janyszek and Mrugala    used   logarithm of the partition function and its second derivative  to  generate  a metric for geometric formulation of thermodynamic as follows \cite{Janyszek}
\begin{equation}
	\,\,\, g_{ij}=  \frac{\partial ^2\ln Z}{\partial \beta_{i} \partial \beta_{j}},\label{eq:Fmetric}
\end{equation}
where $ \beta_{i}=\frac{1}{{{k_B}}}\frac{{\partial S}}{{\partial {X_i}}} $, $ S $ is the entropy and $ {X_i} $ denote the extensive parameters of the thermodynamic systems  (in the following sections we use both Latin and Greek indices for the metric components).  Then, one may calculate other  geometrical quantities such as the scalar and the extrinsic curvatures.  The thermodynamic scalar curvature in a two dimensional  space is determined as follows 
\begin{equation}
	R=\frac{\left|\begin{matrix}g_{{1 1}}&g_{{2 2}}&g_{{1 2}}\\g{_{11,1}}&g{_{22,1}}&g_{{12,1}}\\g_{{11,2}}&g_{{22,2}}&g_{{12,2}}\\\end{matrix}\right|}{2\,\left|\begin{matrix}g_{{11}}&g_{{12}}\\g_{{21}}&g_{{22}}\\\end{matrix}\right|^2},\label{eq:ricci}
\end{equation}
where the parameters of the thermodynamic space are $(\beta_{1},\beta_{2}) $  and ${g_{ij,k}} = \frac{{\partial {g_{ij}}}}{{\partial X_{k} }} $.  It was argued that curvature singularities which are roots of the denominator of $ R $ are corresponded to the second order phase transitions  \cite{Mirza,Fazel}. 
So far, most of attempts in thermodynamic geometry of statistical systems  were in the context of scalar curvature.  However, recently  the  extrinsic curvature was introduced  as a new tool in thermodynamic geometry  \cite{Mansoori}. It was shown that the extrinsic curvature  plays a key role in determining the stability  of the thermodynamic system. Here, we are going to use  extrinsic curvature to study the critical behaviour of some statistical systems.

In the following lines, we review definition of extrinsic curvature in differential geometry. A hypersurface $ \Sigma $  can be defined  by a restriction on coordinates  $\mathcal{H} (X^{a})=0$, where $ X^{a} $ are  coordinates. The extrinsic curvature is expressed as 
\begin{equation}
	K=\mathrm{\nabla}_\mathrm{\mu}\tilde n^\mu=\frac{1}{\sqrt{g}}\partial_\mu\left(\sqrt{ g}\,\tilde n^\mu\right),\label{eq:ext}
\end{equation}
where $ g $ denotes the determinant of the metric and $ \tilde n_\mu$ is a normal vector to hypersurface $\Sigma$. The normalized vector $ \tilde n_\mu $ is given by 
\begin{equation}
	\tilde n_\mu=\frac{\partial_\mu\mathcal{H}}{\sqrt{g^{\mu\nu}\partial_\mu\mathcal{H}\partial_\nu\mathcal{H}}}.\label{eq:vn}
\end{equation}
In this paper, we deal with two dimensional thermodynamic manifolds therefore each hypersurface refers to a one-dimensional curve. In the following, we investigate the extrinsic curvature   of  anyon gas, two dimensional Kagome Ising model and the spherical model. We elaborate on new aspects of using the extrinsic curvature in thermodynamic geometry.
\section{  Anyon gas}\label{sec:anyon}
In this section, our main goal is to derive the nonperturbative thermodynamic extrinsic curvature of an anyon gas and compare its thermodynamic behavior with the corresponding scalar curvature obtained in \cite{anyon1,Mohammadzadeh}.

It is known  that  in $ 3+1 $ dimensions states of identical   bosons  (fermions) are symmetric (antisymmetric) under the  interchange of particles. However, the interchange of two identical particles  in  $ 2+1 $ dimensions may produce    an arbitrary phase $e^{i \pi \alpha}$.  The  statistical parameter $ \alpha $ is in the range  $ 0 \le \alpha  \le 1 $ where, $ \alpha=0 ,1 $ corresponds to bosons and     fermions, respectively. The particles with fractional statistics ($ 0<\alpha <1 $) were called anyons by Wilczek \cite{Wik}.   Over the years, much works have been done to investigate   anyons \cite{ Wik,an1,an2,Wu}. Although fractional exchange statistics mostly defined in two dimensional space,  it may also be appear in $ d=1 $ \cite{d1,Can}. Another type of fractional statistics  is  fractional exclusion statistic that was introduced by Haldane \cite{an1}.  Fractional exclusion statistic   is based on  Hilbert space  and therefore can be defined for an arbitrary dimension $ d\ge 2 $ \cite{27,23,24,25,26}. 

In this section, we will obtain the thermodynamic extrinsic curvature of an anyon gas obeying fractional exclusion statistics. To do this, we  first briefly review  the thermodynamic  formulation of the anyon gas \cite{Huang1,Huang,Huang3}. Wu formulated
the statistical distribution of anyons   by  using Haldane's fraction exclusion statistics as follows \cite{Wu}
\begin{equation}
	{n_i} = \frac{1}{{w(\exp \left[ {\frac{{{\varepsilon _i} - \mu }}{{k_{B}\,T}}} \right]) + \alpha }},
\end{equation}    
where, $ \mu $ denotes the  chemical potential of anyons and $ \varepsilon $ is the energy of the single particle. The function $ w(\chi ) $ fulfills a certain functional equation, $ w(\chi )^{\alpha}[1+w(\chi )^{1-\alpha}]=\chi $ where,  $ w(\chi )= \chi-1 $ for  bosons ($ \alpha=0 $) and for fermions ($ \alpha=1 $), $ w(\chi )= \chi $. In the classical limit ($ \exp \left[ {\frac{{{\varepsilon _i} - \mu }}{{k_{B}\,T}}} \right]\gg 1 $), we have $ w(\chi ) = \chi  + \alpha  - 1 $  therefore the particle number and the internal energy  of anyons  are given by  
\begin{align}
	N &= \sum\limits_i {{n_i}}  = \sum\limits_i {\frac{1}{{\exp [\frac{{{\varepsilon _i} - \mu }}{{k_{B}T}}] + 2\alpha  - 1}}}, \nonumber \\
	U &= \sum\limits_i {{n_i}{\varepsilon _i}}  = \sum\limits_i {\frac{{{\varepsilon _i}}}{{\exp [\frac{{{\varepsilon _i} - \mu }}{{k_{B}T}}] + 2\alpha  - 1}}}
	.\label{eq:adistribut}
\end{align}
The above  particle number and internal energy  reduce to  boson and fermion cases by choosing $ \alpha=0 $ and $ \alpha=1 $, respectively.  It was proposed by  Huang that  a  system of $ N $ particles with  both $\alpha$ fractions of fermions and   $(1-\alpha)$ fractions of bosons  can establish   anyon statistics \cite{Huang}. Therefore,
Based on the Huang's factorized method, the thermodynamic quantity $ Q(\alpha) $ can be written as 
\begin{equation}
	Q_{anyon}= \alpha \, Q_{fermion} + (1-\alpha) \, Q_{boson}.\label{eq:Qanyon}
\end{equation}
As a consequence, the particle number and the internal energy    of the anyons can be written as the combination of the   particle number and the internal energy of fermions and bosons.  Moreover, at finite temperatures for  $ N $ particle number of anyons in the volume $ V $ with a mass $ m $ we have \cite{Wu}
\begin{equation}
	\frac{{{\mu_a }}}{{k_{B}T}} = \frac{{\alpha \,{h^2}}}{{2\pi mk_{B}T}}\frac{N}{V} + \ln \left( {1 - \exp [\frac{{ - {h^2}}}{{2\pi m\,k_{B}T}}\frac{N}{V}]} \right). \label{eq:mukT}
\end{equation}
By rewriting Eq. \eqref{eq:mukT} for boson ($  \alpha=0 $), fermion  ($  \alpha=1 $) and assuming that $ N_{a}= N_{b}=N_{f}=N $ we arrive at
\begin{equation}
	\mu_{a}= \alpha \mu_{f} + (1-\alpha) \mu_{b},\label{eq:mu}
\end{equation}
which is compatible with  the factorized property (Eq.\eqref{eq:Qanyon}),  where  $ a $, $ f $ and $ b $ indices means anyon, fermion and boson, respectively. Therefore, the fugacity of anyons is written as $  z_{a}= z^{\alpha}_{f}  z^{ (1-\alpha)}_{b} $
where
\begin{align}
	z_{b}&=\exp[\frac{\mu_{b}}{ k_{B} T}]= (1-\exp[\frac{- N_{b} \, \beta }{y}], \label{eq:fugacityza}\\ 
	z_{f}&=\exp[\frac{\mu_{f}}{ k_{B} T}]= \exp[\frac{\alpha \, N_{f}\, \beta }{y}] (1-\exp[\frac{- N_{f} \,\beta }{y}] ),\nonumber \\ 
	z_{a} &=\exp[\frac{\mu_{a}}{ k_{B} T}]= \exp[\frac{\alpha \, N_{a} \, \beta }{y}] (1-\exp[\frac{- N_{a}\, \beta }{y}] )
	),\nonumber
\end{align}
where $ \beta =\frac{1}{k_{B} T} $ and $ y=\frac{2\pi m V}{h^2} $.  In the following, we consider  two dimensional space. Therefore, for two dimensional  momentum space we  replace the summation with $ \frac{2\pi m V}{h^{2}} \int\limits_0^\infty  {d\varepsilon }  $  in Eq. \eqref{eq:adistribut}. Then,  the particle number and the internal energy  of anyons are  given by  \cite{Mohammadzadeh} 
\begin{equation}
	\,\,\, N_{a}= y \, \beta^{-1} \,(\alpha \ln(1+z_{f})-(1-\alpha)\ln(1-z_{b})),\nonumber
\end{equation} 
\begin{equation}
	U_{a}= y \, \beta^{-2} \,(-\alpha \, Li_{2}(-z_{f})+(1-\alpha)Li_{2}(z_{b})),\label{eq:UA}
\end{equation} 
where $  Li_{n}(x) $ represents the polylogarithm function. To simplify the relations, we set $ y=1 $. Considering  Eq.\eqref{eq:UA} the parameter  space is $ (\beta,\gamma_{i}) $, where $ \gamma_{i}=-\frac{\mu_{i}}{{k_B}T} $.  The metric elements of boson gas, fermion gas    and anyon gas can be obtained via Eq. \eqref{eq:Fmetric} where $ Z(\beta, \gamma_i)= Tr\, \exp[-\beta H - \gamma_{i} \, N] $ and $  \left\langle H \right\rangle=U  $. Therefore,  using Eq.\eqref{eq:UA}, the metric elements  for the ideal boson gas ($ \alpha=0 $) with the parameters of  $ (\beta,\gamma_{b}) $ is written as  
\begin{equation}
	(g_b)_{{\beta\beta}}=  \frac{\partial ^2\ln(Z_{b})}{ \partial \beta ^2}= - (\frac{\partial U_{b}}{ \partial \beta })_{\gamma_{b}}= 2 \, \beta^{-3} Li_{2} (z_{b}),\nonumber
\end{equation} 
\begin{align}
	\,\,\,	\,\,\,\,\,\,\,\,\,(g_b) _{\beta\gamma_{b}}=   (g_b)_{\gamma_{b}\beta} =\frac{\partial ^2\ln(Z_{b})}{ \partial \gamma_{b}  \partial \beta }= - (\frac{\partial U_{b}}{\partial \gamma_{b} })_{\gamma_{b}}= - \beta^{-2}  \ln({1-z_{b}}),\label{eq:metricb}	 	 
\end{align} 
\begin{equation}
	\,\,\,\,\,\,\,\,\,\,\,\, (g_b)_{\gamma_{b}\gamma_{b}}=  \frac{\partial ^2\ln(Z_{b})}{ \partial \gamma_{f}  \partial \gamma_{f} }=- (\frac{\partial N_{b}}{ \partial \gamma_{b} })_{\beta} =\beta^{-1} \,\frac{z_{b}}{1-z_{b}},\nonumber			
\end{equation} 
and the metric elements of the ideal fermion gas $(\alpha=1) $ with the parameters  $ (\beta,\gamma_{f}) $  are as follows
\begin{equation}
	(g_f)_{{\beta\beta}}=  \frac{\partial ^2\ln(Z_{f})}{ \partial \beta ^2}= - (\frac{\partial U_{f}}{ \partial \beta })_{\gamma_{f}}= - 2\, \beta^{-3} Li_{2} (-z_{f}),\nonumber
\end{equation} 
\begin{align}
	\,\,\,	\,\,\,\,\,\,\,\,\, (g_{f})_{\beta\gamma_{f}}= ( g_{f})_{\gamma_{f}\beta} =\frac{\partial ^2\ln(Z_{f})}{ \partial \gamma_{f}  \partial \beta }= - (\frac{\partial U_{f}}{ \partial \gamma_{f} })_{\gamma_{f}}=  \beta^{-2} \ln({1+z_{f}}), \label{eq:metricf}	 
\end{align} 
\begin{equation}
	\,\,\,\,\,\,\,\,\,\,\,\, (g_f)_{\gamma_{f}\gamma_{f}}=  \frac{\partial ^2\ln(Z_{f})}{ \partial \gamma_{f}  \partial \gamma_{f} }=- (\frac{\partial N_{f}}{ \partial \gamma_{f} })_{\beta} =\beta^{-1} \,\frac{z_{f}}{1+z_{f}}.\nonumber			
\end{equation} 
Then,  the metric elements of anyon takes the following form
\begin{align}
	(g_a)_{\beta\beta}&= \frac{\partial ^2\ln(Z_{a})}{ \partial \beta ^2}= - (\frac{\partial U_{a}}{ \partial \beta })_{\gamma_{a}}\nonumber= 2 \beta^{-3}(-\alpha\, Li_{2}(-z_{f})+(1-\alpha)\,Li_{2}(z_{b})),\\ 
	(g_a)_{\beta\gamma_{a}}=  (g_a)_{\gamma_{a}\beta} &=\frac{\partial ^2\ln(Z_{a})}{ \partial \gamma_{a}  \partial \beta }= - (\frac{\partial N_{a}}{ \partial \beta })_{\gamma_{a}}=  \beta^{-2}(\alpha \ln(1+z_{f})-(1-\alpha)\ln(1-z_{b})),\label{eq:metrica} \\	 	 
	(g_a)_{\gamma_{a}\gamma_{a}}&=  \frac{\partial ^2\ln(Z_{a})}{ \partial \gamma_{a}  \partial \gamma_{a} }=- (\frac{\partial N_{a}}{ \partial \gamma_{a} })_{\beta}= \frac{-1}{(\frac{ \partial \gamma_{a}}{\partial N_{a}})_{\beta}}\nonumber\\&=\beta^{-1}\frac{z_{f}\,z_{b}}{2\alpha z_{f}\,z_{b}+\alpha\, z_{b}-\alpha\, z_{f}- \,z_{f}\,z_{b}+z_{f}}.\nonumber	
\end{align}
Also, It is known that in the  calssical limit,  the equation of state takes the following form \cite{Wu}
\begin{equation}
	PV=Nk_{B}T\, (1+(2\alpha-1)\frac{N \lambda^{2} }{4V}),\label{eq:state}
\end{equation}
where $ \lambda=\frac{h}{{\sqrt {2\pi m{k_B}T} }} $. Therefore,   the interaction is repulsive for $ \alpha>\frac{1}{2} $ and is attractive for $ \alpha<\frac{1}{2} $. For  $ \alpha=\frac{1}{2} $  an ideal equation of state is reached \cite{anyon1}. Nonperturbative thermodynamic geometry of anyon gas  was investigated  in \cite{Mohammadzadeh}.  It was found  that the  scalar curvature has  opposite signs for boson  $ (R>0) $ and fermion $ (R<0) $ cases \cite{anyon1,Janyszek}. It is known that the positive scalar curvature indicates the attractive interaction while the negative one leads to repulsive interaction \cite{anyon1,Rupplist1}.  Moreover,  at the Bose-Einstein condensation the corresponding thermodynamic scalar curvature diverges.  It was shown that anyons  behave as an ideal classical gas  at $ \alpha =\frac{1}{2} $ where $ R=0 $ \cite{anyon1}. For a thorough discussion, the interested reader is referred to \cite{Mohammadzadeh}.

Now,  we will compute the thermodynamic extrinsic curvature of bosons, fermions and anyon  gas.   We assume    a constant hypersurface, $ \beta=cte $, in order  to   compare  the results with thermodynamic properties obtained via thermodynamic scalar curvature  for  an isotherm in \cite{Mohammadzadeh}.
The  starting point  is to compute a normal vector of  the hypersurface. So, using  Eq.\eqref{eq:vn} and Eq.\eqref{eq:metricb} the normal vector $ (\tilde n_b)_\beta  $ for bosons is  given by
\begin{align} 
	(\tilde n_b )_\beta &= \frac{{{\partial _\beta }\mathcal{H} }}{{\sqrt {{(g_b)^{\beta \beta \,}}{\partial _\beta }\mathcal{H} \,{\partial _\beta }\mathcal{H} } }} \\ \nonumber &= \frac{1}{{\sqrt {{(g_b)^{\beta \beta \,\,}}} }}  = \,\left({\frac{{2{z_b}\,L{i_2}({z_b}) + ({z_b} - 1)\ln {{(1 - {z_b})}^2}}}{{{\beta ^3}{z_b}}}}\right)^{\frac{1}{2}} ,
\end{align}
it should be noted that another component   ($ (\tilde n_b )_{\gamma_{b}} $)  is equal to zero. For fermions  using the same calculations we have
\begin{align}
	(\tilde n_f )_\beta =\frac{1}{{\sqrt {{(g_f)^{\beta \beta \,}}} }}  =\left({ - \frac{{2{{z_f}}\,L{i_2}( - {{z_f}}) + ({{z_f}} + 1)\,{{\ln }}({{z_f}} + 1)}^2}{{{\beta ^3}{{z_f}}}}}\right)^{\frac{1}{2}} ,
\end{align}
where we have used  Eq.\eqref{eq:metricf}, and the other  component  ($ (\tilde n_f )_{\gamma_{f}}  $)   is equal to zero. The normal vector with upper indices has both components ($ (\tilde n_i) ^\beta $  and $ (\tilde n_i) ^{\gamma_{i}} $ )  due to  the off-diagonal metrics (Eqs.\eqref{eq:metricb}, \eqref{eq:metricf},  \eqref{eq:metrica}). Therefore, for bosons we arrive at
\begin{align}  
	(g_b)^{{\beta\beta}} (\tilde n_b )_\beta&=(\tilde n_b) ^\beta =\left({\frac{\beta ^3 {z_b}}{2 {z_b}\, L{i_2}({z_b})+({z_b}-1) \ln (1-{z_b})^2}}\right)^{\frac{1}{2}},\\ \nonumber\\ \nonumber
	(g_b)^{{\beta\gamma_{b}}} (\tilde n_b )_{\beta}&=(\tilde n_b) ^{\gamma_{b}}= \frac{{(1-{{z_b}})\ln (1 - {z_b})}}{{{z_b}}}\\ \nonumber&\times\left({\frac{{\beta \,{z_b}}}{{2{z_b}\,L{i_2}({z_b}) + ({z_b} - 1)\,{{\ln }}(1 - {z_b})^2}}}\right)^{\frac{1}{2}} , 
\end{align}
and for fermions we have
\begin{align}  
	(g_f)^{{\beta\beta}} (\tilde n_f )_\beta&=	(\tilde n_f) ^\beta  =\left({-\frac{\beta ^3{z_f}}{2 {z_f}\,L{i_2}(-{z_f})+({z_f}+1) \ln ({z_f}+1)^2}}\right)^{\frac{1}{2}},\\\nonumber\\ \nonumber
	(g_f)^{{\beta\gamma_{f}}} (\tilde n_f )_{\beta}&=	(\tilde n_f) ^{\gamma_{f}}=- \frac{{({{z_f}} + 1)\ln ({{z_f}} + 1)}}{{{{z_f}}}}\\ \nonumber&\times\left({ - \frac{{\beta \,{z_f} }}{{\left( {1 + {z_f}} \right)\ln {{\left( {1 + {z_f}} \right)}^2} + 2{z_f}\, L{i_2}({-z_f})}}}\right)^{\frac{1}{2}}.  \label{eq:normalf}	
\end{align}
As a result,   the extrinsic curvature for bosons takes the following form
\begin{align}  
	K_b&=\frac{1}{\sqrt{g_b}}\partial_\mu \left(\sqrt{ g_b}\,n_b^\mu \right) \\ \nonumber
	&=\frac{1}{\sqrt{g_b}} \left(\partial _{\beta} (\sqrt{ g_b}  \, (\tilde n_b)^\beta )+\partial_{\gamma_{b}}(\sqrt{ g_b} \,(\tilde n_b)^{\gamma_{b}}) \right),\label{eq:Kbb}	
\end{align}
where $ g_b $ denotes the determinant of the  metric for bosons  and $ \partial_\mu = (\frac{\partial}{\partial \beta},\frac{\partial}{\partial \gamma_b})$. The computation for fermions is  the same  and therefore,  the extrinsic curvature for bosons and fermions  can be found as follows
\begin{align}
	{K_{b}}{\rm{ }} &= \frac{{({z_b} + \ln (1 - {z_b}))}}{{2{z_b}}}\left({\frac{{\beta\, {z_b}}}{{2{z_b}\,L{i_2}({z_b}) + ({z_b} - 1)\ln {{(1 - {z_b})}^2}}}}\right)^{\frac{1}{2}}  ,\\ \nonumber
	{K_{f}}{\rm{ }} &= \frac{{({z_f} - \ln (1 + {z_f}))}}{{2{z_f}}}\left({ - \frac{{\beta\, {z_f}}}{{2{z_f}\,L{i_2}( - {z_f}) + ({z_f} + 1)\ln {{(1 + {z_f})}^2}}}}\right)^{\frac{1}{2}}.
\end{align}
\begin{figure}[t]
	\centering
	\includegraphics[width=.8\columnwidth]{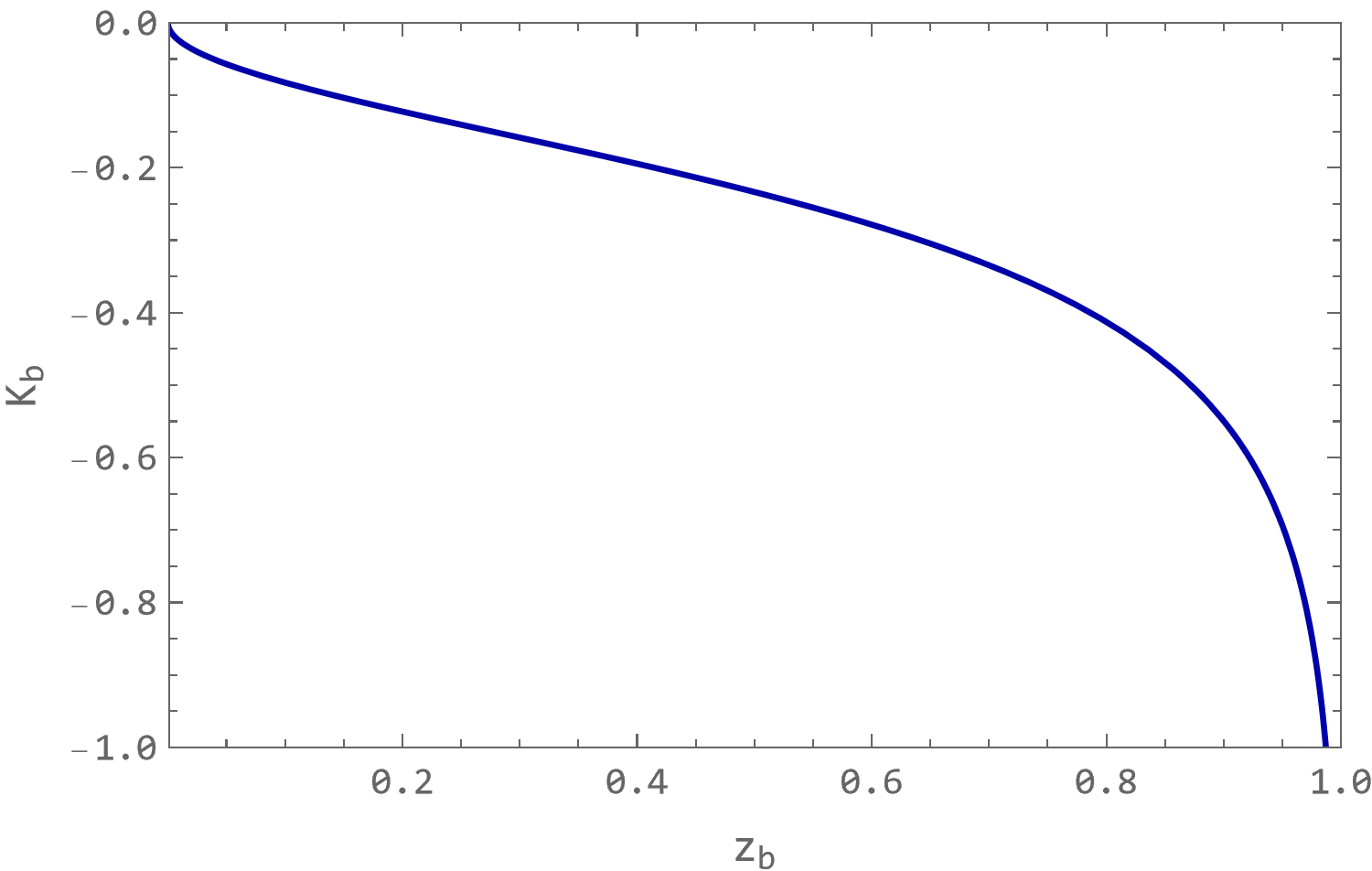}
	\caption{ \normalfont  The thermodynamic  extrinsic curvature for bosons as a function of $ z_{b} $. The singular point at $ {z_b}=1 $ displays Bose-Einstein condensation phase transition. We have set $\beta=1$.}
	\label{fig:extrinsicb}	
\end{figure}
\begin{figure}[t]
	\centering
	\includegraphics[width=0.8\columnwidth]{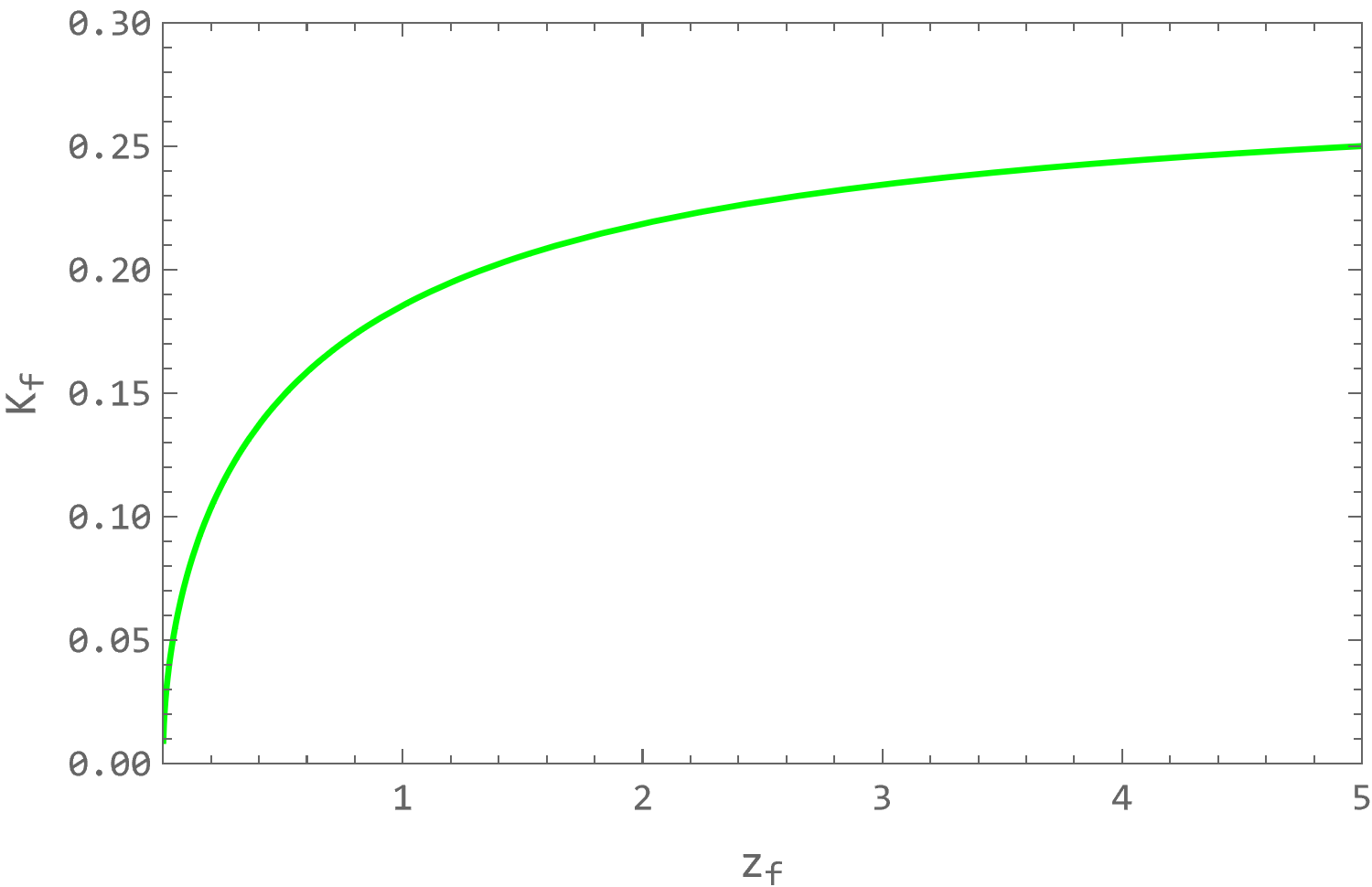}
	\caption{ \normalfont The thermodynamic  extrinsic curvature for fermions as a function of $ z_{f} $  and  $\beta=1$.}
	\label{fig:extrinsicf}	
\end{figure}
Fig.\ref{fig:extrinsicb} and Fig.\ref{fig:extrinsicf}  show the  extrinsic curvatures of bosons and fermions,  respectively, with respect to  fugacities  $ z_{b} $ and $ z_{f} $. 
It is seen  that the extrinsic curvature has opposite signs for bosons and fermions. Moreover, There is a singularity for bosons which is located at $ z_{b}=1 $. This point in higher dimensions corresponds to the Bose-Einstein condensation phase transition \cite{Janyszek}. It is known that  boson gas are less stable than fermion gas \cite{Mohammadzadeh,Janyszek}. Therefore, we observe the   thermodynamic extrinsic curvature of bosons and fermions behaves properly as a geometric tool in  study of statistical mechanics of the anyon gas. 

Now, let us investigate the  extrinsic curvature of  anyons. The parameters of space for anyons is $ (\beta,\gamma_{a}) $, where $ \gamma_{a}=-\frac{\mu_{a}}{{k_B}T} $. Now, by using Eq.\eqref{eq:fugacityza}, we can replace $   {z_f}$ by $  \frac{{z_b}}{1-{z_b}} $ in the metric components (Eq.\eqref{eq:metrica}). The unit normal vectors of anyons with upper indicies take the following form
\small
\begin{flalign} 
	&\begin{aligned}
		(\tilde n_a)^\beta &=\left({\frac{\beta ^3 {z_b}}{2 (1-\alpha ) {z_b}\, L{i_2}({z_b})-2 \alpha\,  {z_b}\, L{i_2}\,\left(\frac{{z_b}}{{z_b}-1}\right)+((1-\alpha)\, {z_b}-1) \left(\alpha  \ln \left(\frac{1}{1-{z_b}}\right)-(1-\alpha) \ln (1-{z_b})\right)^2}}\right)^{\frac{1}{2}},\\ \\
		(\tilde n_a)^{\gamma_a} &= \frac{{((1-\alpha ){{z_b}} - 1)\left( {\alpha \ln \left( {\frac{1}{{1 - {{z_b}}}}} \right) + (\alpha  - 1)\ln (1 - {{z_b}})} \right)}}{{{{z_b}}}}  \\&\times \left({  \frac{{\beta\, {{z_b}}}}{{ 2(   1-\alpha){{z_b\,L}}{{{i}}_2}({{z_b}})-2\alpha\, {{z_b\,L}}\,{{{i}}_2}\left( {\frac{{{{z_b}}}}{{{{z_b}} - 1}}} \right) + ((1-\alpha )\,{{z_b}} - 1){{\left( {\alpha \ln \left( {\frac{1}{{1 - {{z_b}}}}} \right) + (\alpha  - 1)\ln (1 - {{z_b}})} \right)}^2}}}}\right)^{\frac{1}{2}}.
	\end{aligned}&&
\end{flalign}
\normalsize
The extrinsic curvature of anyons is given by
\begin{align}  
	K_{a}&=	\frac{1}{\sqrt{g_a}} \left(\partial _{\beta} (\sqrt{ g_a}  \, (\tilde n_a)^\beta )+\partial_{\gamma_{a}}(\sqrt{ g_a} \,(\tilde n_a)^{\gamma_{a}})\right ),\label{eq:kAn}
\end{align}
where $ g_a $ denotes the determinant of the metric  of anyons which components are given  in  Eq.\eqref{eq:metrica}. Eq.\eqref{eq:mu}  implies       $ {\gamma _a} = \alpha \,{\gamma _f} + (1 - \alpha ){\gamma _b} $,  therefore we   can write the second term  in  Eq.\eqref{eq:kAn} as follows
\begin{align}
	\partial_{\gamma_{a}}\left(\sqrt{ g_a} \, (\tilde n_a)^{\gamma_{a}}\right) & =\frac{1}{{\frac{{\partial {\gamma _a}}}{{\partial \left( {\sqrt {{g_a}} {{\left( {{{\tilde n}_a}} \right)}^{{\gamma _a}}}} \right)}}}}\\ \nonumber &={\left( {\frac{{\alpha  }}{{\partial {\gamma _f}(\sqrt {{g_a}}  {{({{\tilde n}_a})}^{{\gamma _a}}})}} + \frac{{(1 - \alpha )}}{{\partial {\gamma _b}(\sqrt {{g_a}}  {{({{\tilde n}_a})}^{{\gamma _a}}})}}} \right)^{ - 1}}\\ \nonumber &=	{\left( {\frac{{\alpha (1 + \frac{{{z_b}}}{{1 - {z_b}}}) }}{{\partial {\gamma _b}(\sqrt {{g_a}}  {{({{\tilde n}_a})}^{{\gamma _a}}})}} + \frac{{(1 - \alpha )}}{{\partial {\gamma _b}(\sqrt {{g_a}}  {{({{\tilde n}_a})}^{{\gamma _a}}})}}} \right)^{ - 1}} ,\label{eq:kAA}
\end{align}
where we have used  Eq.\eqref{eq:fugacityza},    $\frac{\partial}{\partial {\gamma_{f}}}=\frac{\partial{\gamma_{b}}}{\partial {\gamma_{f}}} \frac{\partial}{\partial {\gamma_{b}}}$ and $\frac{\partial{\gamma_{b}}}{\partial {\gamma_{f}}}= (1+\frac{z_b}{1-z_b}) ^{-1} $.   
\begin{figure}[t] 
	\centering
	\includegraphics[width=0.8\columnwidth]{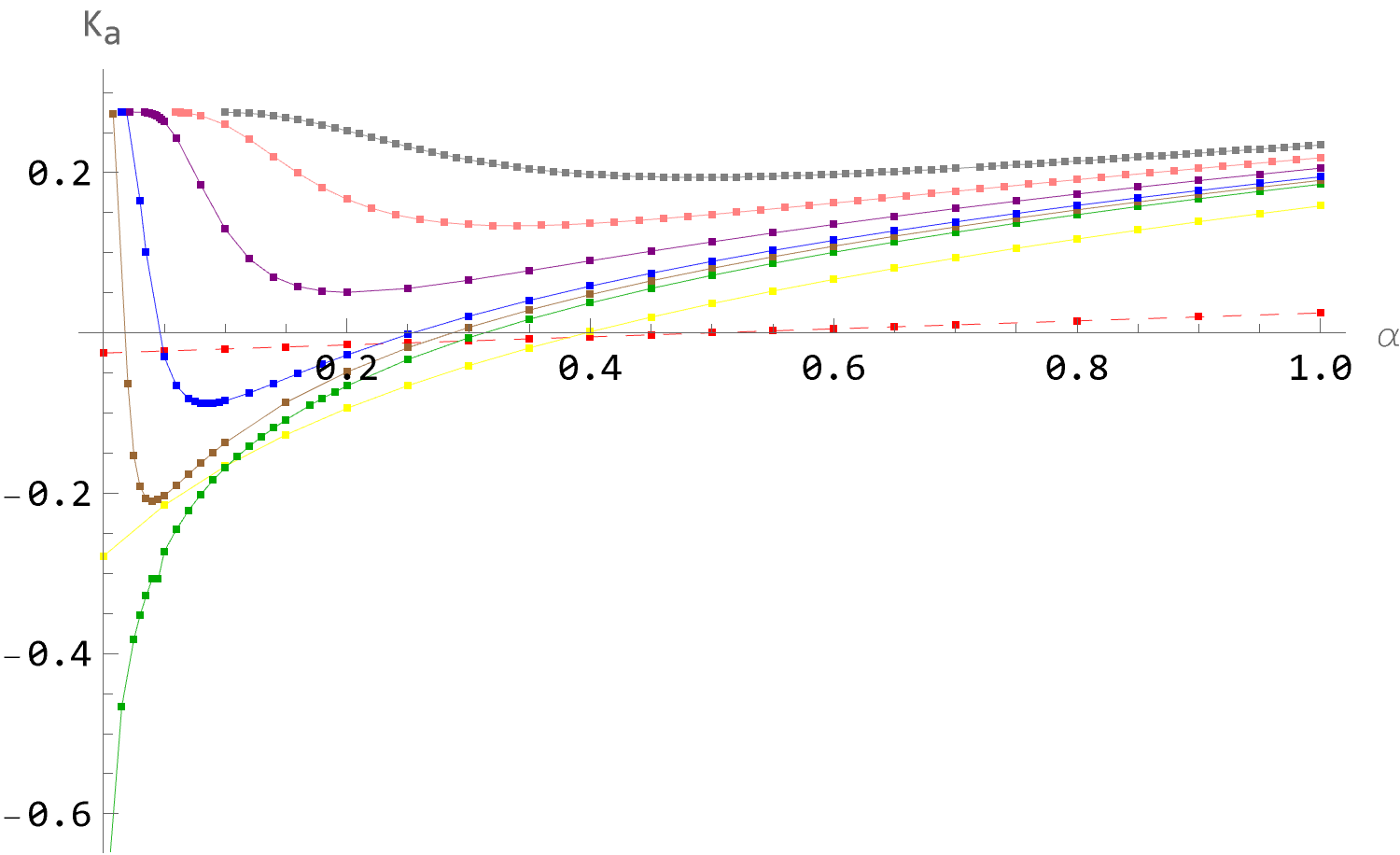}   
	\caption{ \normalfont  The thermodynamic extrinsic curvature of  anyon gas as a function of $ \alpha $. The values of anyon fugacity have been considered as $ z_{a}= 0.01 $ (dashed red  curve),   $ 0.6 $ (yellow curve), $ 1.0 $ ( green curve), $1.1 $ (brown), $ 1.2 $ (blue), $ 1.5 $ (purple),  $ 2.0 $ (pink) and $ 3.0 $ (gray upper curve). We have set $\beta=1$ for all diagrams. }\label{fig:coex1}
\end{figure}
Fig.\ref{fig:coex1} shows the extrinsic curvature   for different  values of anyon fugacities. In the classical limit we have obtained Eq.\eqref{eq:state} therefore at  $ \alpha= \frac{1}{2}$  it represents a classical ideal gas  ($ PV=N k_{B}T $). It is seen for $ z_{a}=0.01 $  the extrinsic curvature behaves as an ideal classical gas whose  sign  changes at $ \alpha= \frac{1}{2}$. For larger values of $ z_{a} $,    we observe a fermionic behavior where the extrinsic curvature is positive.  The extrinsic curvature has a minimum point  for $  z_{a}>1 $ that exists in the range $  \alpha<  \frac{1}{2}$.   As the value of anyon fugacity increases the   minimum point goes toward the bigger values of $ \alpha  $.  Furthermore, for some values of anyon fugacities ($  z_{a}>1$), there are two points of $ \alpha $ where  the extrinsic curvature has the same amount, indicating  a duality relation between those points.   For $  z_{a}=1 $,   the extrinsic curvature goes to negative infinity (green curve) at $ \alpha= 0 $ that represents the bosonic behavior. The results are in agreement with behavior  of thermodynamic Ricci scalar that was investigated in  \cite{Mohammadzadeh}.
\begin{figure}[t]
	\centering
	\subfloat[  \label{subfig:kaa}]{\includegraphics[width=0.7\columnwidth]{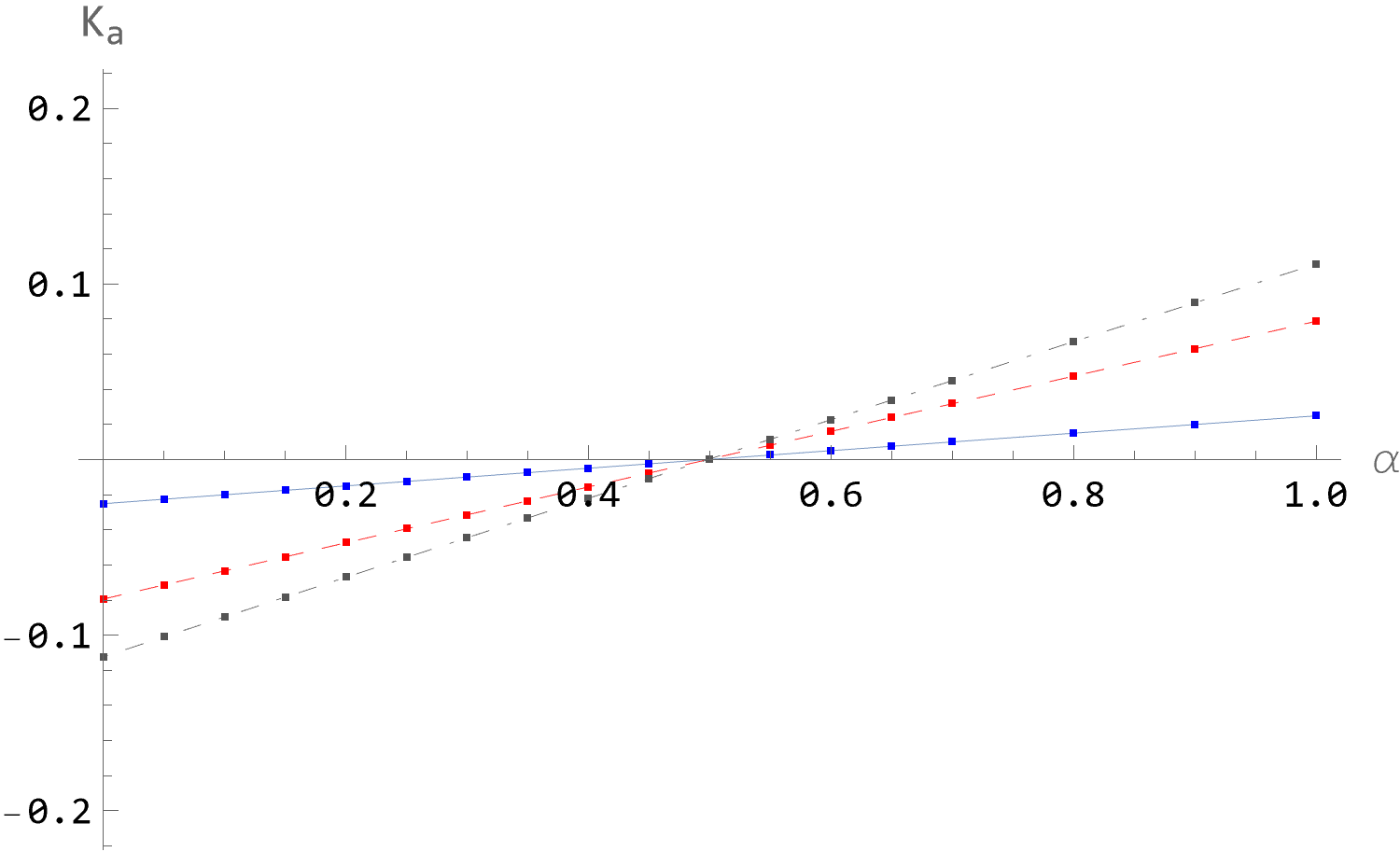}}
	
	\subfloat[\label{subfig:kab}]{\includegraphics[width=0.7\columnwidth]{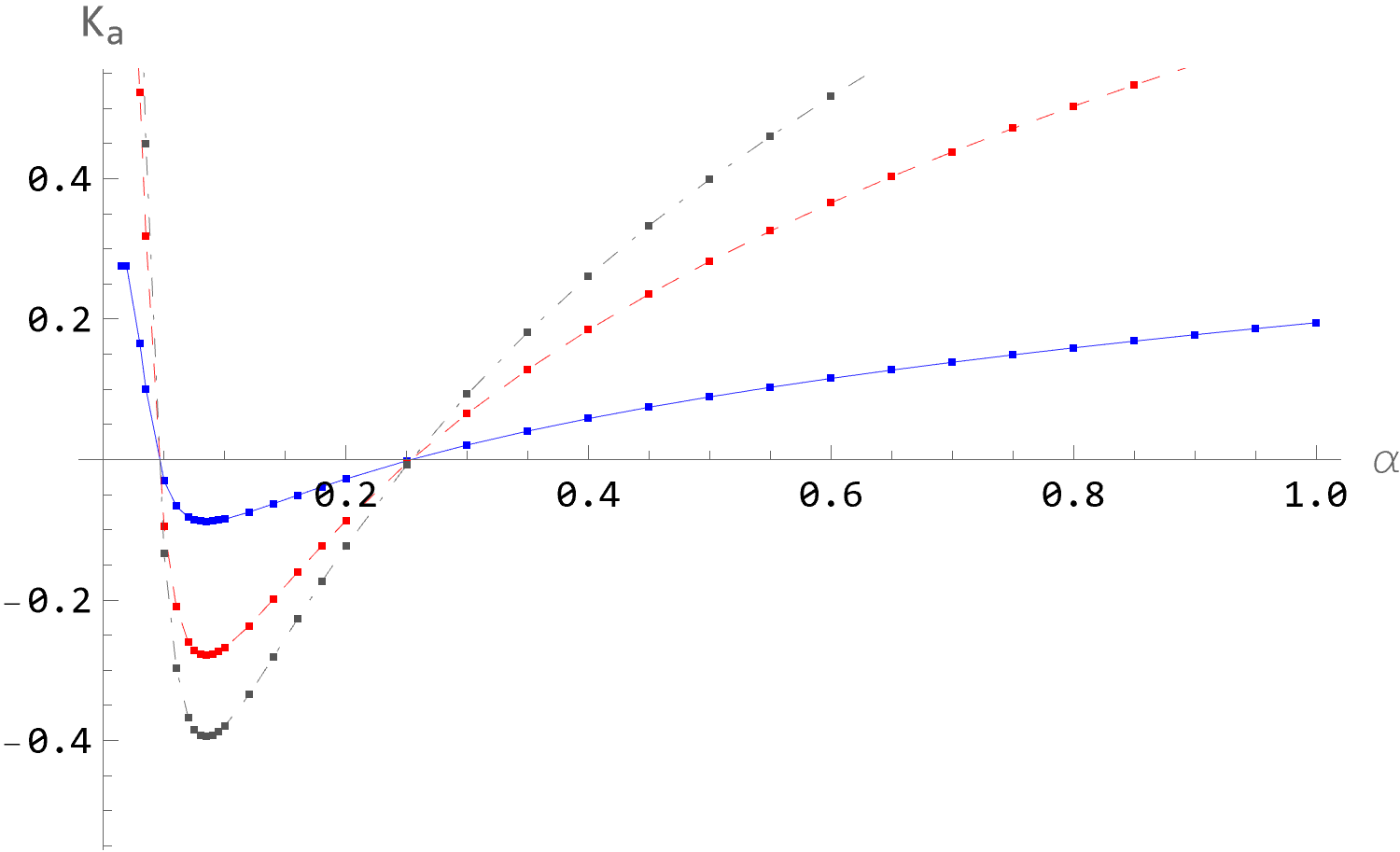}}
	
	\caption{\normalfont The thermodynamic  extrinsic curvature of  anyon gas as a function of $ \alpha $ for different $ \beta $ hypersurfaces. The diagrams are for $ \beta=20 $ (dotted-dashed gray curve), $  \beta=10 $ (dashed-red curve) and  $ \beta=1 $ ( solid blue curve) at  constant anyon fugacities   (a) $ z_{a}=0.01 $ (b) $ z_{a}=1.2 $.}
	\label{fig:extrinsickaaa}
\end{figure}

Now, let us investigate the thermodynamic extrinsic curvature for different values of $ \beta $ hypersurfaces.    Fig.\ref{fig:extrinsickaaa}      shows  the thermodynamic extrinsic curvature as a function of $ \alpha $ for three  values of inverse temperatures $ \beta $. First, we consider the classical limit and $ z_{a}=0.01 $.   There is a fixed point that the value of the thermodynamic extrinsic curvature is equal to zero for all values of inverse temperatures, where $ \alpha=0.5 $ in  Fig.\subref* {subfig:kaa}. At this fixed point, the gas  particles behave as  semions and do not have  bosonic or fermionic behavior. Then, we restrict our attention to the other interesting  case $  z_{a}=1.2 $.  Fig.\subref* {subfig:kab}     shows by increasing the value of inverse temperature $ \beta $ (upper curves to downer curves) the minimum  absolute value of  the extrinsic curvature increases which moves  toward the small values of $ \alpha $. Therefore, for larger values of $ \beta$ or as $ T \to 0$, the thermodynamic extrinsic curvature  for $ \alpha=0 $  has the bosonic behavior with $ K_{a}<0 $. Also,  we observe semionic behavior of the  anyon gas particles   where $ \alpha=0.25 $. Therefore, for values away  from the classical limit, the fixed point with $ K_{a}=0 $ moves toward the small values of $ \alpha $. Also,  we   find  two different points  with $ K_{a}=0 $. These new fixed  points can not be  obtained  by using the    thermodynamic  Ricci scalar curvature.  
\section{  Critical systems  }\label{sec:critic}
Thermodynamic  geometry of the  two dimensional Kagome Ising model was investigated in  \cite{kagome}.   It was shown  that at zero magnetic  filed the thermodynamic   scalar  curvature behaves as  $ R  \sim \varepsilon^{\alpha-1} $ where  $ \varepsilon= \beta-  \beta_{c}$  and $ \alpha=0 $ ($\beta_{c}$ denotes the critical point). For positive values of $ \alpha $  the critical behavior of the thermodynamic scalar curvature is  $ R\sim \varepsilon^{\alpha-2} $  \cite{spheric}. 
The critical scaling behavior of the  thermodynamic scalar curvature not only determines  the phase transition points   but  also can be used   to  extract the critical exponent $\alpha  $.  Consider the critical region, where we expect a standard scaling form of the free energy per site  as follows
\begin{equation}
	f(\varepsilon,h)= \lambda ^{-1} 	f(\varepsilon \lambda^{a_{\varepsilon}} ,h \lambda ^{a_{h}} ),
\end{equation}
where  $ a_{h}  $ and  $ a_{\varepsilon} $  are scaling dimensions of spin operators and the energy, respectively.  In the high temperature region ($ \varepsilon= \beta_{c}-\beta  >0$), using the scaling assumptions  and  the scaling function  $ \psi_{+} $  we arrive at
\begin{equation}
	f(\varepsilon,h)= \varepsilon ^{\frac{1}{a_{ \varepsilon}}} 	\psi_{+}( h \,\varepsilon ^{-\frac{a_{h}}{a_{\varepsilon}}}).\label{eq:sc}
\end{equation}
In the second order transition, the scalar curvature can be written as $  R\sim \xi ^{d}$, where $ d $ is the dimension of the system and $ \xi $  is the correlation length.  Here, we are going to obtain  the critical exponents via the scaling behavior of the extrinsic curvature. We study  the two dimensional Kagome Ising model and the spherical model.  The relevant  parameters  of the spin   systems are  $ (\beta, h) $ where $ \beta=\frac{1}{k_{B}T} $ and $ h $ is the external magnetic field. 

\subsection{   Kagome Ising model  }\label{sec:criticK}
\begin{figure}
	\centering
	\includegraphics[width=.5\textwidth]{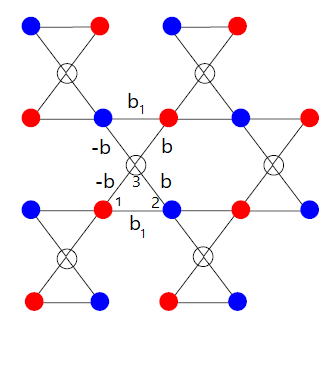}
	\caption{ \normalfont Kagome Ising lattice in a magnetic field. Colorful circles show  magnetic Ising spins $ 1,2 $ and open circles non-magnetic spins $ 3 $. }.	\label{fig:2dkagome}	
\end{figure}

\begin{figure}[t]
	\centering 
	\includegraphics[width=0.8\columnwidth]{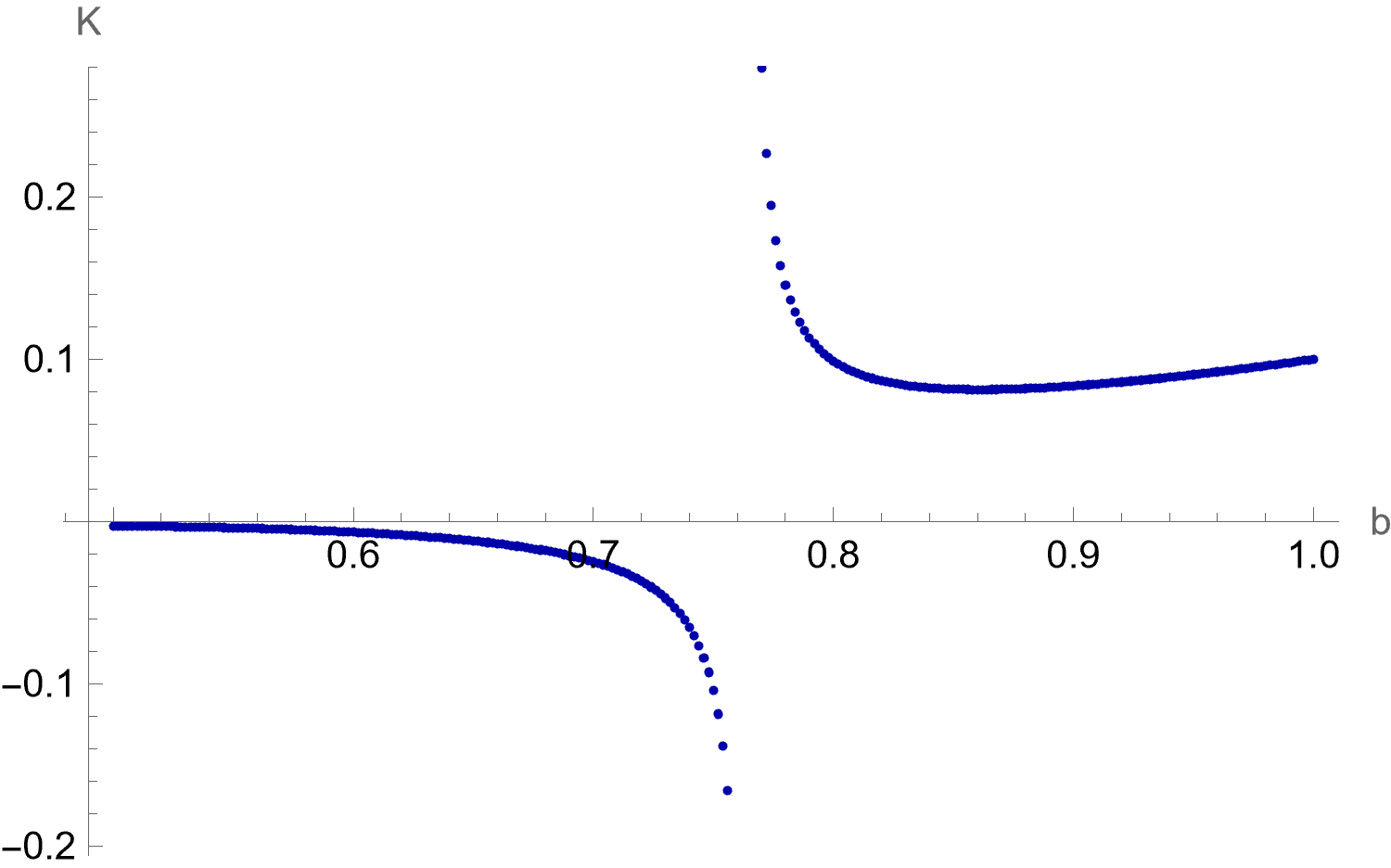}
	\caption{\normalfont Extrinsic  curvature of two dimensional Kagome Ising model  at $ h =0 $ hypersurface, the critical point is $ b_{c}=0.76 $. }
	\label{fig:hextrinsic}
\end{figure}

The  two dimensional Ising model in a magnetic field  background has been studied by different methods \cite{Baxter1966,Wuu}.  Fisher  obtained  a unique exact solution of an antiferromagnetic Ising model \cite{ fisher}. The  Fisher model was defined on a square lattice in the nonzero magnetic field. It was also generalized to the Ising model on the Kagome lattice and  in a  magnetic field background which is soluble for a few cases \cite{Baxter,Azaria,Lin,Lu}. In one of the cases, the Kagome Ising model  was solved by  using the equivalency between those partition functions of the Kagome   and the Honycomb Ising models \cite{Lu}. That case transforms into the Fisher model for the square lattice. In recent years, Kagome lattice has  received attention  for its applications   in  high-Tc superconductivity. Here, we obtain the thermodynamic extrinsic curvature and the critical exponent using the Fisher model. For an Ising model on the Kagome lattice, see Fig.\ref{fig:2dkagome}, the interacting energy of every triangle composed  of the spins $ \sigma _1,  \sigma _2, \sigma _3  $  is given by 
\begin{equation}
	-J_{1}\sigma_{1}\sigma_{2}+J(\sigma_{2}\sigma_{3}-\sigma_{3}\sigma_{1})-\mu H(\sigma_{1}+\sigma_{2}),
\end{equation}
where $ H $ denotes the magnetic field which affects  the $ \frac{2}{3} $ of sites on the lattice.
The reduced  field  determines by $h=\beta H$ and reduce interactions   are  $b=\beta J ,b_{1}=\beta J_{1}$.   The  Ising spin $ \sigma _3$  with zero magnetic moment does not couple to the magnetic field. Therefore,   $ \sigma _1,  \sigma _2$  have a super-exchange interaction
with the  spin $\sigma _3 $ (non-magnetic).  The Fisher model can be obtained from this model by assuming $  b_{1}=0 $. The partition functions of the  Ising model on the kagome and Honeycomb lattices are equivalent at zero magnetic field, $ Z_{KG} = F^{N} Z_{HC} $, where  $ N $ represents the number of lattice sites in the  Honeycomb model and the magnetic spins in the Kagome lattice,  the parameter  $ F $ is a constant. 
The per-site free energy for Honeycomb lattice  and the per magnetic spin free energy  are, respectively, given by
\begin{align}
	f_{\text{HC}} = \lim _{N\rightarrow\infty} N^{-1} \ln Z_{{\text{HC} }},\  \  \ \text{and}  \   \  \ \
	f_{\text{KG} }= \lim _{N\rightarrow\infty} N^{-1} \ln Z_{{\text{KG} }},  \label{eq:hony}
\end{align}
The relation of  $ F_{\text{HC}} $ was reported in \cite{fhc2,fhc1}. 
Therefore, by using Eq.\eqref{eq:hony} we arrive at the free energy of Kagome Ising model
\begin{align}
	f=\frac{1}{16\pi^{2}} \int_{0}^{2\pi}\ \int_{0}^{2\pi}\ d\theta d\varphi \  \xi  (\theta,\varphi)+\frac{3}{4} \ln 2+\ln F  ,
\end{align}
where
\begin{align}
	\xi (\theta,\varphi)&=\cosh 2r_{1} \cosh^{2}2r 
	-\sinh^{2}2r \cos(\theta+\varphi)
	-\sinh 2r_{1} \nonumber\\ &\times\sinh 2r (\cos\theta+\cos\varphi) +1,
\end{align}
and $ r_{1} $ and $ r $ are, respectively, as  a function of $ b_{1} $ and $ b $. We have the Fisher model in the case  $ b_{1}=0 $.  Therefore, by Setting $ b_{1}=0 $ in the above relation,  we can write the  free energy per spin  as follows \cite{Lu}
\begin{equation}
	f=f_{1}(b,h)+ f_{2}(r),
\end{equation}
where $ f_{1}(b,h), f_{2}(r) $ are defined  as follows
\begin{align}
	f_{1}(b,h)&=\frac{1}{4}\ln\{\cosh^2h\ ({\sinh^2h+\cosh^22b})\}+\frac{3}{2}\ln2
	,\\\nonumber\\
	f_{2}(r)&=\frac{1}{16\pi^2}\int_{0}^{2\pi} \int_{0}^{2\pi}d\theta \, d\varphi \ln\{\cosh^22r
	-\sinh2r\ (\cos\theta+\cos\varphi)\}\nonumber.
\end{align}
Also,  $ r $  fulfills the below relation
\begin{equation}
	r= -  \frac{1}{4} \ln(\frac{\sinh^2h+\cosh^22b}{\cosh^2h}).
\end{equation}
With some simplifications $f_2(r)  $ takes the following form \cite{kagome}
\begin{equation}
	f_2(r)= \frac{1}{\pi}\int_{0}^{\frac{\pi}{2}}d\theta
	\, \ln(\frac{1}{2}(1+\sqrt{1-\kappa^2\sin^2\theta}) )+\frac{1}{2}\ln(\cosh^22r),
\end{equation}
where $\kappa=-\frac{2\sinh2r}{\cosh^22r}$. Here, the parameters of the thermodynamic space are $ (b,h) $ and  from the metric defined in Eq.\eqref{eq:Fmetric} where  $ \ln Z=f$, the  metric elements of two dimensional  Kagome Ising model are given by 
\begin{equation}
	g_{ij} \equiv f_{ij}=\partial_{i}\partial_{j}f.
\end{equation}
Therefore,  we arrive at
\begin{align}
	g_{bb} \equiv f_{bb}	&= \frac{\partial f_{b}}{\partial  b},\nonumber \\  
	g_{hh} \equiv f_{hh}	&= \frac{\partial f_{h}}{\partial  h},\label{eq:kggmetric}\\   
	g_{hb} =g_{bh} \equiv f_{bh}	&= \frac{\partial f_{h}}{\partial  b},\nonumber
\end{align}
where
\begin{align}
	f_{b} & \equiv \frac{\partial f}{\partial b}=\frac{\partial f_{1}}{\partial b} + \frac{\partial  f_{2}}{\partial r}\frac{\partial r}{\partial b}, \nonumber \\ 
	f_{h} & \equiv \frac{\partial f}{\partial h}=\frac{\partial f_{1}}{\partial h} + \frac{\partial f_{2}}{\partial r}\frac{\partial r}{\partial h}\label{eq:kgf}. 
\end{align}
The relations  are  lengthy so   we write   them  in the general form. The first derivatives of    free energy  can be found in  Appendix.   The  thermodynamic geometry of the Kagome Ising model in two dimensions  was studied  in \cite{kagome}. The results show  the scalar curvature $ R $ is positive  at high temperature that  shows  disordered state (paramagnetic), while it is negative for low temperature ordered state (antiferromagnetic). Moreover, as the temperature decreases
the scalar curvature diverges at plus infinity. It was shown the critical line where the scalar curvature also  diverges at the same points coincides with  the Fisher expression
\begin{equation}
	h=\text{arccosh}\ (\sqrt{\frac{\sqrt{2}-1}{2}}\ \text{sinh}\ (2b)),\label{eq:cline}
\end{equation}
and in the zero field case ($ h=0 $) upon setting the denominator of $ R $ equals to zero,    the critical point is obtained as
\begin{align}
	b_{c}=\text{arccosh} (\sqrt{\frac{2+\sqrt{2}}{2}}) ,\label{eq:criticb}
\end{align}
where $ b_{c} $  denotes the critical point.
Now, we will obtain the extrinsic curvature using the Fisher model. We first  assume a constant  $ b $ hypersurface,    therefore from Eq.\eqref{eq:vn} we have $ \tilde n_{b}=\frac{{{\partial _b }\mathcal{H} }}{{\sqrt {{g^{bb}}{\partial _b}\mathcal{H} \,{\partial _b }\mathcal{H} } }}  $. From this relation, the upper indices unit normal vectors for constant $ b $ hypersurface  takes the following form 
\begin{align}
	\tilde n^{b}&= g^{bb}\, \tilde n_{b}=\frac{f^{bb}}{\sqrt{f^{bb}}},\\
	\tilde n^{h}&= g^{bh} \tilde n_{b}=\frac{f^{bh}}{\sqrt{f^{bb}}}\nonumber,
\end{align}
\begin{figure}[t]
	\centering
	{\includegraphics[width=0.8\columnwidth]{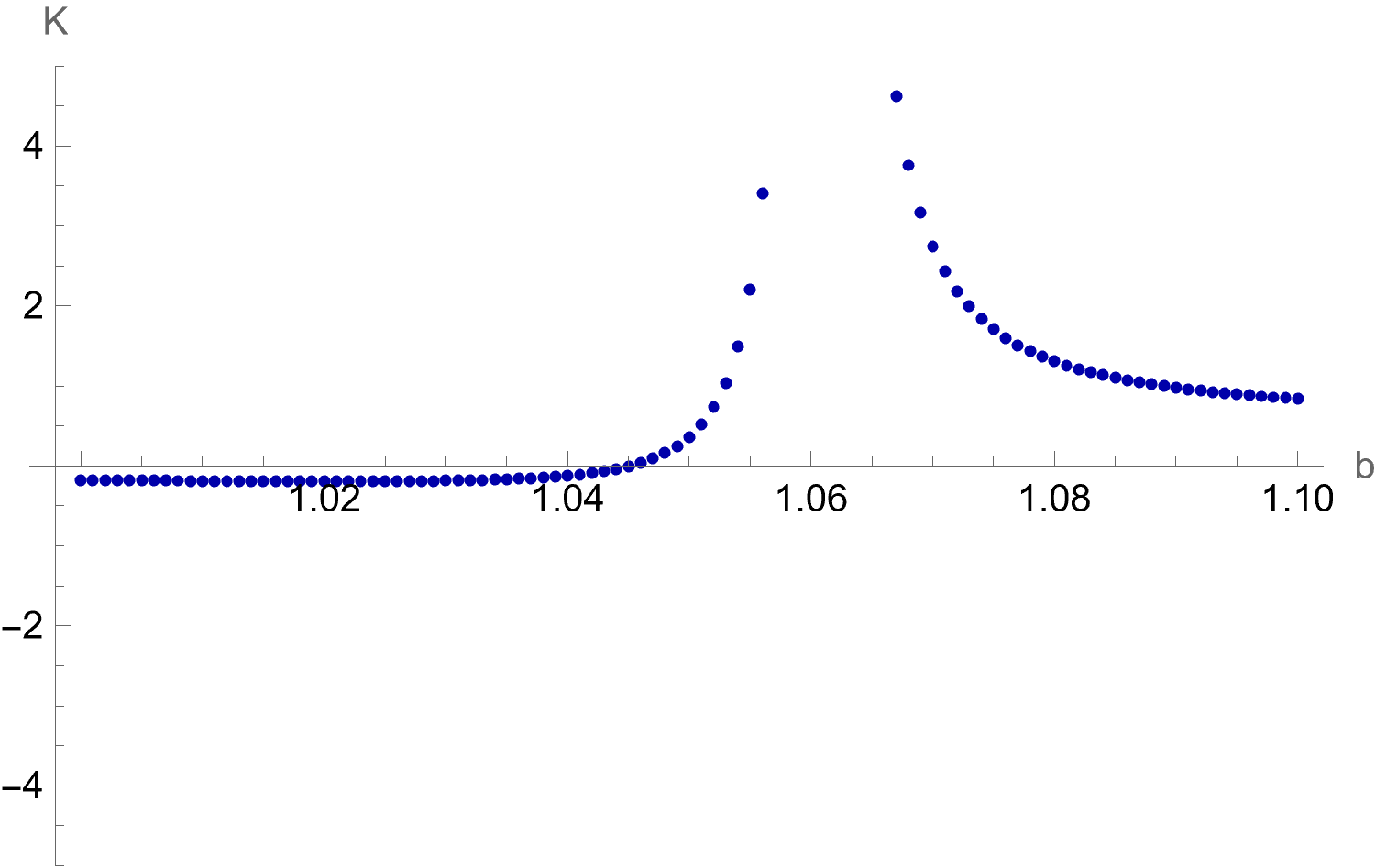}}

	\caption{\normalfont Thermodynamic extrinsic  curvature $ K $ of two dimensional Kagome Ising model for $ h $ hypersurface. At   $  h=1.24 $, the thermodynamic extrinsic curvature $K  $ diverges at the critical point $  b=1.06 $. }
	\label{fig:extrinsick&ta}
\end{figure}
\begin{figure}[t]
	\centering
	
	{\includegraphics[width=0.8\columnwidth]{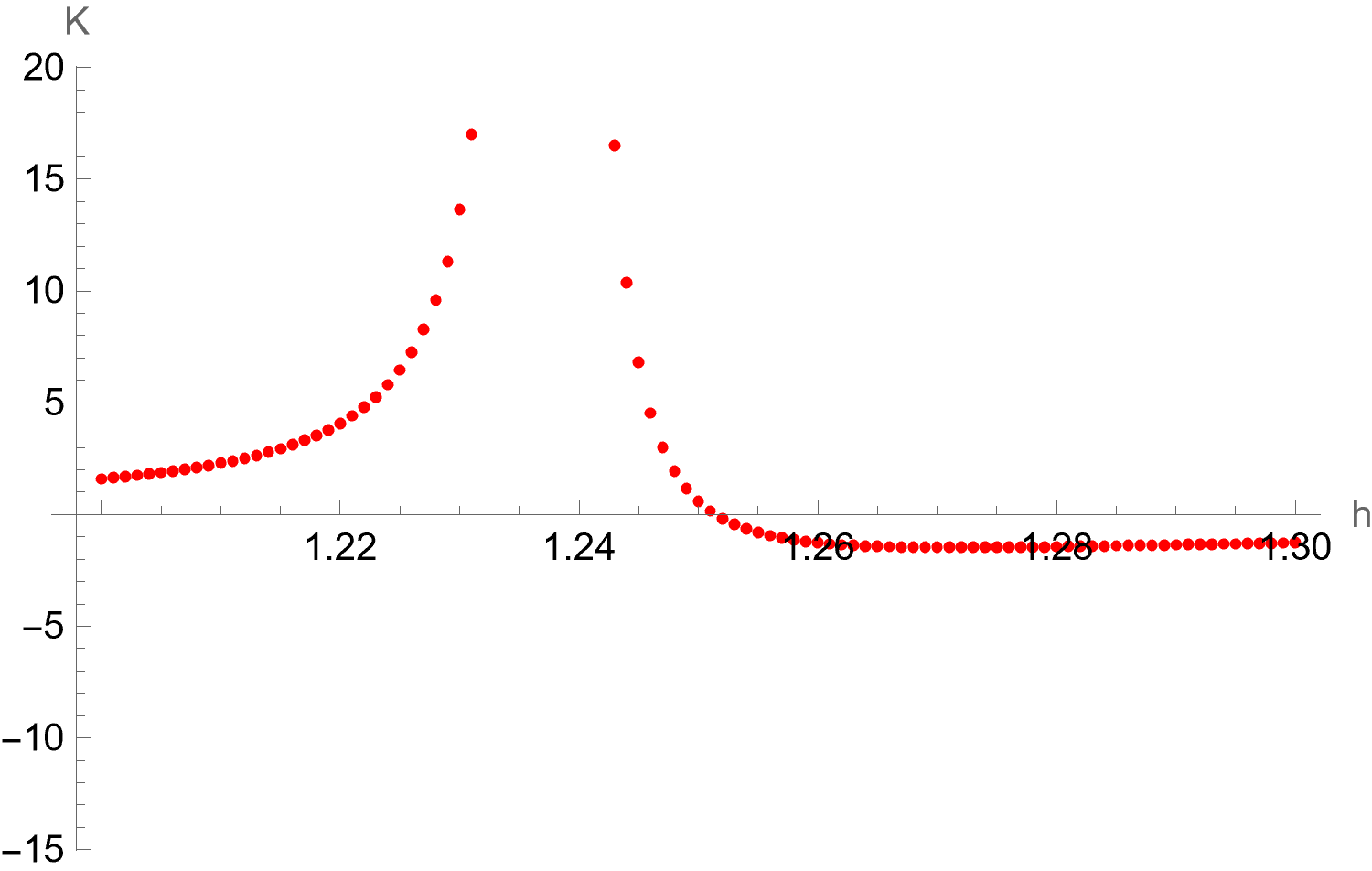}}
	
	\caption{\normalfont  Thermodynamic extrinsic curvature $ K $ of two dimensional Kagome Ising model for $ b $ hypersurface.   At  $ b=1.06 $, $ K $ diverges at the critical point $   h=1.24 $.  }
	\label{fig:extrinsick&tb}
\end{figure}
where we have used  Eq.\eqref{eq:kggmetric}. Then, by making  use of Eq.\eqref{eq:ext} the extrinsic curvature is expressed as  
\begin{align}  
	K=\frac{1}{\sqrt{g}} (\partial _{b} \left(\sqrt{ g}  \, \tilde n^b \right)+\partial_{h}\left(\sqrt{ g} \,\tilde n^{h} \right) ),\label{eq:extrinsick}
\end{align}
where $ g $ denotes the determinant of the  two dimensional Kagome Ising metric.  Also, the computation of the thermodynamic  extrinsic curvature for the  $h$ constant  hypersurface  can be done by using the following  unit normal vectors ($ \tilde n_{h}=\frac{1}{\sqrt{g^{hh}}}  $)
\begin{align}
	\tilde n^{b}&= g^{bb}\, \tilde n_{h}=\frac{f^{bb}}{\sqrt{f^{hh}}},\\
	\tilde n^{h}&= g^{bh} \tilde n_{h}=\frac{f^{bh}}{\sqrt{f^{hh}}}.\nonumber
\end{align}
We   have depicted the thermodynamic extrinsic curvature of the 2-d Kagome Ising model in Fig.\ref{fig:extrinsick&ta} and  Fig.\ref{fig:extrinsick&tb}, respectively,  for different $ h $ and $ b $ hypersurfaces. The points $ h=1.24 $ and $ b=1.06 $ satisfy the  Fisher expression in Eq.\eqref{eq:cline}. 
Fig.\ref{fig:extrinsick&ta} and Fig.\ref{fig:extrinsick&tb}  show  divergence points of the thermodynamic extrinsic  curvature, properly.  The divergence points   correspond to the  phase transition points.    Moreover, Fig.\ref{fig:extrinsick&ta} shows that $ K <0 $ at high temperature  or small value of $ \beta $ while $ K >0 $ for law temperature. Therefore, by decreasing  the temperature  a transition occurs from  the disordered (paramagnetic) to an  ordered state (antiferromagnetic).  In Fig.\ref{fig:extrinsick&tb}, by increasing  an external field we  observe a phase transition  from the ordered state ($ K>0 $) to the disordered state ($ K<0 $) at the critical point $ h_{c}=1.24 $. We also observe that   signs of the thermodynamic extrinsic curvature $ K $ and the  thermodynamic scalar curvature  are opposite to each other.
The thermodynamic extrinsic curvature  is depicted at  zero magnetic field in Fig.\ref{fig:hextrinsic}. It is seen that the divergence point  corresponds to  the critical point in Eq.\eqref{eq:criticb} exactly. The correspondence of   the   divergence points of the thermodynamic extrinsic and scalar curvatures is an important issue that should be studied more in the feature.

\begin{figure}[t]
	\centering
	\includegraphics[width=0.8\columnwidth]{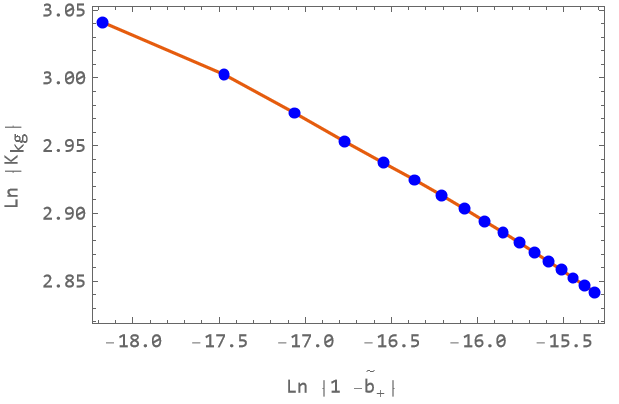}
	\caption{ \normalfont The numerical values (blue points) of $ \ln \left| K_{kg} \right| \ $ as a function of $\ln(1-\tilde{b}_+)$ for two dimensional Kagome Ising model near the critical point $ b_{c}=0.76428 $. The data are fitted by the red line with the slope $ 0.07148 $.}	\label{fig:extrinsip}	
\end{figure}

Now, we are going to derive the critical behavior of the thermodynamic extrinsic curvature. As usual, we expect a power law behavior near critical points. Therefore,  around  the critical points,  the extrinsic curvature  behaves  as $ K\propto \, - {\left( {1 - {{\tilde b}_ + }} \right)^a} $ or
\begin{equation}
	\ln|K|=-a\ln\left(1-\tilde{b}_+ \right)+ c,\label{eq:lnr}
\end{equation}
where  the reduced parameter   $\tilde{b}_+ =\frac{b}{b_{c}}  $   and  $ {b_{c}} $ is the critical point, where $ K $ diverges. In the following lines, we obtain the scaling  behavior of the extrinsic curvature at zero magnetic field, $ h=0 $ hypersurface.  Using the definition of the extrinsic curvature   in Eq.\eqref{eq:extrinsick}, we  can compute Eq.\eqref{eq:lnr} for different values of $ b $ near the critical point $ b_{c}  $ (Eq.\eqref{eq:criticb}).   We extrapolate the numerical  values   and   find the coefficients $ a $ and $ c $ by a numerical method. We  indicate the numerical   results 
in Fig.\ref{fig:extrinsip}   by blue points  and the  red  fitting line.  From Fig.\ref{fig:extrinsip} we obtain $ a\simeq  0.07148 $ and $ c\simeq 1.75139 $.  Extrapolating the numerical  values, our results are consistent with $ a=0 $. So,   we find that the thermodynamic extrinsic curvature behaves as $ K\sim(b- b_{c} )^{\alpha}\equiv \varepsilon^{\alpha} $, where $ \alpha=0 $ as  it was expected. However, it was found that the thermodynamic  scaling behavior of the  scalar curvature is $ R\sim \varepsilon^{\alpha-1} $ with $ \alpha=0 $ which  is the same as 2d-Ising model on planar random graph with $ \alpha=-1 $. 
There is a unit of difference between the power of the thermodynamic scalar and extrinsic curvatures. Also, we have recently studied thermodynamic of    pure Lovelock black holes by using thermodynamic geometry  method \cite {ebrahimi}. We have found the scaling  behavior of the scalar and the extrinsic curvatures are, respectively, $ R\sim \varepsilon^{\alpha-2} $  and  $ K\sim \varepsilon^{\alpha-1} $ with $ \alpha=0 $.  It is interesting that a universal behavior appears as   $ \frac{R}{K}  \sim  \frac{1}{\varepsilon}$  for different models such as pure Lovelock black holes and Kagome Ising model. 
\newpage
\subsection{ The spherical model  }\label{sec:critics}
The spherical model is a generalized form of  the two dimensional Ising model.  Kac  introduced  this model that  spin variable $ s_{i} $   have  arbitrary values \cite {Kac} such as 
\begin{equation}
	\sum\limits_i {{s_i}^2}  = N.
\end{equation}
Therefore, the value of the spin can vary continuously on the sphere with the radius $ N^{\frac{1}{2}} $ from  $ -N^{\frac{1}{2}}$ to   $ N^{\frac{1}{2}}$. 
This approach is useful for  spin models with an external field. In 1952, Berlin and Kac solved this model for one, two and three dimensions. Three dimensional spherical model presents transition at finite temperature while one and two dimensional cases do not that differs  from the two dimensional Ising model.  The  model is reliable for entire temperatures. It was shown that the critical exponents of the Ising model on  two dimensional planar random graphs are the same as the three dimensional spherical model  \cite{random,inform} and $ \alpha=-1 $. As $ \alpha $ is negative  the scalar curvature behaves as $ R\sim \varepsilon ^{\alpha-1}  $ rather than $ R\sim \varepsilon ^{\alpha-2}  $. Moreover, for  $ d\ge 4 $ the model can be described by  a mean field theory with $\alpha=0  $.  The spherical model's  partition function  is given by 
\begin{equation}
	Z=	\int {d{s_1}}  \ldots d{s_N}\exp \,(\beta \sum\limits_{ < ij > } {{s_i}{s_j}}  + h\sum\limits_i {{s_i}} )\,\delta (\sum\limits_i {{s_i}^2}  - N), \label{eq:smodel}
\end{equation}
where $ {s_i} $ denotes the value of a spin,  $ N $ is the  number of sites and $ h $ is the magnetic field.   Using the saddle-point   method and  $ f=\ln Z$, the  free energy  in the thermodynamic limit ($ N\to \infty  $) is written as
\begin{equation}
	f = \frac{1}{2}Log\left( {\frac{\pi }{\beta }} \right) + \beta \,z - \frac{1}{2}g(z) + \frac{{{h^2}}}{{4 \beta \,(z - d)}}, \label{eq:free}
\end{equation}
where
\begin{equation}
	g(z)=\frac{1}{{{{(2\pi )}^d}}}\int\limits_0^{2\pi } {d{\omega _1} \ldots d{\omega _d}} \, Log\left( {z - \sum\limits_{k = 1}^d {cos({\omega _k})} } \right). \label{eq:gz}
\end{equation}
Therefore, the saddle point for the free energy  is given by
\begin{equation}
	g'(z) = 2\beta  - \frac{{{h^2}}}{{2\beta {{(z - d)}^2}}}.\label{eq:gzz}
\end{equation}		
We focus our attention  to the case $ d=3 $,	 since there is no phase transition for $ d=1 $	and $ d=2 $. Using Eq.\eqref{eq:gzz} with $ h=0 $ we have
\begin{equation}
	\frac{{dz}}{{d\beta }} = \frac{2}{{g''(z)}},\label{eq:dz}
\end{equation}			
therefore
\begin{equation}
	\frac{{{d^2}z}}{{d{\beta ^2}}} = \frac{{ - 4{g'''(z)}}}{{g''{{(z)}^3}}}.\label{eq:d2z}
\end{equation}
The critical point   is  given by $ z=d=3 $ for  $ h=0 $ and the phase transition correlate  with the  spontaneous magnetization in three dimensional lattice \cite {Kac}. To reach $ g(z) $ in the critical region, we  derive the second derivative of $ g(z) $ by differentiating Eq.\eqref{eq:gz} and expanding around   small  values of $ \omega_k $   therefore we have
\begin{equation}
	g''(z) \sim \frac{{ - 1}}{{2\sqrt 2 \pi }}{\left( {z - 3} \right)^{\frac{{ - 1}}{2}}}\label{eq:g2},
\end{equation}
and
\begin{equation}
	g'''(z) \sim \frac{{ 1}}{{4\sqrt 2 \pi }}{\left( {z - 3} \right)^{\frac{{ - 3}}{2}}}.\label{eq:g3}
\end{equation}
Therefore by integrating we have
\begin{equation}
	g'(z) \sim \frac{{ 1}}{{4\sqrt 2 \pi }}{\left( {z - 3} \right)^{\frac{{ 1}}{2}}}+g'(3),
\end{equation}
Now, inserting $ h=0 $ in Eq.\eqref{eq:gzz} and using the above relation we arrive at the following relation
\begin{equation}
	\left( {z - 3} \right) \sim 8{\pi ^2}{(\beta  - {\beta _c})^2} \sim {\varepsilon ^2},
\end{equation}
where $ \beta _c=\frac{g'(3)}{2}\approx 0.25, $. Therefore by using Eq.\eqref{eq:g2} and Eq.\eqref{eq:g3}, the scaling  behavior of $\frac{{{dz}}}{{{d\beta }}}  $ and $\frac{{{{d^2}z}}}{{{d{\beta ^2}}}}  $ from Eq.\eqref{eq:dz} and Eq.\eqref{eq:d2z}  may be written as
\begin{align}
	\mathop {\lim }\limits_{z \to 3} \frac{{dz}}{{d\beta }} &= \mathop {\lim }\limits_{z \to 3} \left\{ { - 4\sqrt 2 \pi {{\left( {z - 3} \right)}^{\frac{1}{2}}}} \right\} = 0, \\
	\mathop {\lim }\limits_{z \to 3} \frac{{{d^2}z}}{{d{\beta ^2}}} &= 16{\pi ^2}.
\end{align}
It was shown  that for $ \alpha=-1 $ the scalar  scalar curvature behaves as $ R\sim \varepsilon^{\alpha-1} $  \cite{spheric22}. We expect that this type of scaling behavior of $ R $ to be universal for  $ \alpha<0 $.   In the next section, we derive the   scaling behavior of the thermodynamic  extrinsic curvature by a simple method.

\section{ The  critical  behavior of the extrinsic curvature}\label{sec:standard}
In this section, we derive the scaling form of the extrinsic curvature by using  Eq.\eqref{eq:sc}.   We also  simplify  the notation and define $ A=\frac{1}{a_{\varepsilon}} $, $ C=\frac{-a_{h}}{a_{\varepsilon}} $.  With respect to standard critical exponents  we have $ A=2-\alpha $ and $ A+C=\beta $ \cite{spheric22}.  Therefore,  from Eq.\eqref{eq:sc}  the scaling form of the free energy   is given by  $f= \varepsilon ^A 	\psi_{+}( h \,\varepsilon ^{C}) $, where $ \varepsilon= \beta_{c}-\beta $ and $ h $ is the external magnetic field. The scaling form of the metric elements with parameters ($ \beta , h $)   are calculated in the following form
\begin{align}
	g_{\beta \beta } &= \partial _\beta ^2f = A(A - 1)\,\phi (0),\label{eq:scal1} \\ \nonumber
	g_{hh} &= \partial _h^2f = {\varepsilon ^{A + 2C}}\psi_{+} ''(h \,\varepsilon ^{C} ), \\
	g_{\beta h} &= g_{h\beta } =\partial _\beta \partial _h f =-\varepsilon  ^{A+C-1} \left((A+C)\, \psi_{+} '\left(h\, \varepsilon ^C\right)+ C h \,\varepsilon ^C \psi_{+} ''(h \,\varepsilon ^{C} )\right) ,\nonumber
\end{align}
which  is for $ \alpha<0 $ ($ A>2 $). In this case, we expect that  the  specific heat is a  constant  at the critical point and we define it by  $ A(A - 1)\,\phi (0) $, where $ \phi (0) $ is a constant function. We assume a constant $ \beta $ hypersurface. Using  Eq.\eqref{eq:vn}, the unit normal vector  is given by
\begin{align}
	{{\tilde n} 	_\beta} & = \frac{1}{{\sqrt {g^{{\beta \beta }}} }}\\ \nonumber &=(\frac{A (A-1)\, \phi(0) \, \psi_{+} ''\left(h\, \varepsilon^C\right)-\varepsilon^{A-2} \left((A+C) \,\psi_{+} '\left(h\, \varepsilon^C\right)+C\, h \,\varepsilon^C \psi_{+} ''\left(h\, \varepsilon^C\right)\right)^2}{\psi_{+} ''\left(h\, \varepsilon^C\right)})^{\frac{1}{2}}.
\end{align}
The upper indices of the vector are given by
\begin{align}
	{\tilde n^{\beta} }&= {g^{{\beta \beta }}}\, {{\tilde n} 	_\beta}=\frac{{g^{{\beta \beta }}}}{{\sqrt {g^{{\beta \beta }}} }},\label{eq:negative}\\
	{\tilde n^{h} }&= g^{\beta h}  {{\tilde n} 	_\beta}=\frac{g^{\beta h}}{\sqrt{{g^{{\beta \beta }}}}}.\nonumber 
\end{align} 
\begin{table}[t]
	\centering
	\caption {\normalfont The standard scaling behavior of the thermodynamic scalar and extrinsic curvatures  for the spherical model at the critical point ($ h=0 $), where, $ \alpha<0 $   and $ \alpha>0 $ are, respectively,  for  $  d=3$ and $ d=4 $.  Derivation of the  scaling thermodynamic extrinsic curvature was done on  an isotherm  hypersurface. }\label{tab:1}
	\begin{tabular}{ccc}
		\hline 
		Extrinsic curvature  & Scalar curvature &\mbox{$\alpha$}   \\
		\hline 
		$ K\sim\varepsilon^{\alpha}$      &  $R \sim\varepsilon^{\alpha-1}$  &\mbox{$\alpha<0 $ } \,\,    \\ 		
		$ K\sim\varepsilon ^{{\frac{1}{2}} (\alpha-2)} $   &  $R\sim \varepsilon ^{ \alpha-2} $ &	\mbox{$\alpha>0 $ }     \\
		\hline 
	\end{tabular}
	
\end{table}
Then the extrinsic curvature by using Eq.\eqref{eq:ext}   at zero magnetic field ($ h=0 $) near the critical point ($ \varepsilon  \to 0  $) is  written in the following form
\begin{align}
	K&=\frac{1}{{\sqrt {g}}}\left({\partial _\beta } (\sqrt {g} \,{\tilde n^{\beta} })+{\partial _h }(\sqrt {g} \,{\tilde n^{h} })\right)\nonumber \\
	&= \frac{\varepsilon \left(\psi_{+} '''(0)\, (A+C) \,\psi_{+} '(0)-(A+2 C)\, \psi_{+} ''(0)^2\right) \left((A-1) A \,\phi(0) -\frac{(A+C)^2 \psi_{+} '(0)^2 \,\varepsilon^{A-2}}{\psi_{+} ''(0)}\right)^{\frac{1}{2}}}{2 \psi_{+} ''(0) \,{\left((A-1) A\, \phi(0) \, \varepsilon^2 \,\psi_{+} ''(0)-(A+C)^2 \,\psi_{+} '(0)^2\, \varepsilon^A\right)}} \nonumber \\
	&=-\frac{A+2 C}{2\, \varepsilon(A (A-1)\, \phi(0))^{\frac{1}{2}} },
\end{align}
where, $ g $ denotes the determinant of the metric in Eq.\eqref{eq:scal1}.  We have used  $ \psi '(0)=0 $ and $ \psi ^{'''}(0)=0 $ since the odd $ h $ derivatives of the scaling function   should  vanish at $ h=0 $ for Ising-like models  \cite{yyans}.    
Therefore in the case $ \alpha<0 $, we find the scaling behavior of the extrinsic curvature is $ K\sim  \varepsilon^{-1} $ which is consistent with $ K\sim  \varepsilon^{\alpha} $, where $ \alpha=-1 $. However, the scaling behavior of the  thermodynamic  scalar curvature can be  obtained as $ R \sim {\varepsilon}^{\alpha-1 } $ for  $ \alpha<0 $ which is the same as the two dimensional Ising model on planar random graph.  This result is in accordance with the standard
scaling behavior of the thermodynamic scalar curvature for negative $ \alpha $. Therefore  we expect that the extrinsic curvature  of three dimensional spherical  model behaves as $ K\sim \varepsilon^{\alpha} $. 

The spherical model has mean field behavior in $ d\ge 4 $   with    $ \alpha=0 $. The scaling  behavior of the thermodynamic scalar curvature  is  $ R \sim {\varepsilon}^{-2 } $ which corresponds to  $ R \sim {\varepsilon}^{\alpha-2 } $, where $ \alpha=0 $.  In this case, the scaling metric elements are given by
\begin{align}
	g_{\beta \beta }&= \partial _\beta ^2f,\label{eq:metric2} \\ \nonumber
	{g_{hh}} &= \partial _h^2f = {\varepsilon ^{A + 2C}}\psi_{+} ''(h\, \varepsilon ^{C} ), \\
	{g_{\beta h}} & = g_{h\beta } =\partial _\beta \partial _h f =-\varepsilon  ^{A+C-1} \left((A+C)\, \psi_{+} '\left(h \,\varepsilon ^C\right)+ C\, h \,\varepsilon ^C \psi_{+} ''(h\, \varepsilon ^{C} )\right),\nonumber
\end{align}
where  $ f= \varepsilon ^A 	\psi_{+}( h \,\varepsilon ^{C}) $ as before. Therefore,  $ {g_{\beta \beta }} $ in Eq.\eqref{eq:metric2}  is not constant  and  is given by $  A(A-1){\varepsilon ^{A -2}}\,\psi_{+} (0  ) $  at $ h=0 $ which is different from $ {g_{\beta \beta }} $ in Eq.\eqref{eq:scal1}.
The normal vector components  indices  $ {\tilde n_{\beta} }  $, $ {\tilde n^{\beta} }  $ and ${\tilde n^{h} }  $  can be obtained as before. We derive the standard   scaling behavior of the thermodynamic  extrinsic curvature   at  zero magnetic field ($ h=0  $) near the critical point ($ \varepsilon \to 0 $) as follows
\begin{align}
	K&=\frac{1}{{\sqrt g}}\left({\partial _\beta } (\sqrt {g} \,{\tilde n^{\beta} })+{\partial _h }(\sqrt {g} \,{\tilde n^{h} })\right) \nonumber\\ &=\frac{\left((A+C)\,\psi_{+} '(0)\, \psi_{+} '''(0)  -(A+2 C)\, \psi_{+} ''(0)^2\right)   ({\frac{\psi ''(0)\, \varepsilon^{2-A}}{(A-1) A \,\psi_{+} (0)\, \psi_{+} ''(0)-(A+C)^2 \,\psi_{+} '(0)^2}})^{\frac{1}{2}}}{2 \varepsilon\, \psi_{+} ''(0)^2} \nonumber\\
	&=-(A+2 C) ({\frac{\varepsilon^{-A}}{(A-1) A\, \psi_{+} (0)}})^{\frac{1}{2}},
\end{align} 
where  we have used $ \psi '(0)=0 $ and  $ \psi ^{'''}(0)=0 $. Therefore, we expect that the scaling behavior of the thermodynamic extrinsic curvature of the spherical model behaves as  $  K \sim\varepsilon ^{{\frac{1}{2}} (\alpha-2)}$ for $ d\ge4 $     while  the scaling behavior of the scalar curvature    is $ R\sim \varepsilon ^{\alpha-2} $, where $ \alpha=0 $. It should be noted that we obtain the same scaling behavior of the thermodynamic extrinsic curvature by  assuming a constant $ h $ hypersurface. Moreover,  we expect that the same behavior to be valid  for $ \alpha>0  $ in $ d\ge4 $, we have gathered the whole results in Table \ref{tab:1}.
\section{Conclusions}\label{sec:conclud}
The thermodynamic extrinsic curvature is a geometric tool for observing some aspects of statistical mechanics. It is a new geometric window to observe the critical phenomena. In this paper, we investigated the  thermodynamic extrinsic curvature of an anyon gas. We explored nonperturbative thermodynamic extrinsic curvature of an anyon gas $ K_{a} $  at constant fugacities $ z_{a}=0.01 $ and $ z_{a}=1.2 $ and obtained certain fixed points that particles behave as  semions. This is a new result that can not be obtained  in the context of the thermodynamic Ricci scalar curvature.  Then, we studied the two dimensional Kagome Ising model and showed that      $ K\sim \varepsilon^{\alpha} $  with $ \alpha=0 $.  Therefore, for two dimensional Ising models with $ \alpha \le 0 $   we  expect  that the scaling of the   extrinsic curvatures behave as  $ K\sim \varepsilon^{\alpha} $. Then,   we  derived the scaling behavior of the extrinsic curvature  related to the spherical model. We found that the  scaling extrinsic curvature  behaves as  $ K\sim \varepsilon^{\alpha} $ with $ \alpha=-1  $.   We also studied the spherical model in  $ d\ge 4 $ and obtained  $  K \sim\varepsilon ^{{\frac{1}{2}} (\alpha-2)} $, where $  \alpha=0 $.     However, for  $ \alpha> 0 $ we  derived a different  scaling behavior for the extrinsic curvature  as $ K \sim\varepsilon ^{{\frac{1}{2}} (\alpha-2)}$. In this work, we obtained some important properties of the extrinsic curvature. The  fermion gas and an ordered state of the two dimensional Kagome Ising model have positive thermodynamic extrinsic curvatures.  It should be noted  that the sign of the  thermodynamic extrinsic curvature is positive for  stable states.  Also, we showed the results from  the computing  thermodynamic extrinsic curvature of Kagome Ising model  $ K\sim\varepsilon^{\alpha}$  is consistent   with what we expected from  the standard scaling behavior  of the thrmodynamic extrinsic curvature. We hope to study the extrinsic curvature and related geometric pictures for other physical systems in the near future.
\section*{Acknowledgments}
We would like to thank Isfahan   University of Technology for financial support.

\appendix
\section*{Appendix } \label{App:A}
Usefull equations that are   related to Eq.\eqref{eq:kgf}:
\begin{align}
	\frac{\partial f_{1}}{\partial b} = \frac{\sinh2b\,\cosh2b}{\sinh^{2}h+\cosh^{2}2b} , \ \ \ \ \ \ \ \frac{\partial f_{1}}{\partial h }= \frac{\tanh h\,(\cosh^{2}2b+\cosh^{2}h)}{\cosh4b+\cosh2h},\label{eq:A.1} \tag{A.1}
\end{align}
and 
\begin{align}
	\frac{\partial f_{2}}{\partial b}&=\frac{1}{16\pi(\cosh4b+\cosh2h)^{2}}\,
	\text{csch}\,[\frac{1}{2}\log(\cosh^{2}2b \, \text{sech}^{2}h+\tanh^{2}h)]\,
	\nonumber \\  &\times \text{sech}\,[\frac{1}{2}\log(\cosh^{2}2b \, \text{sech}^{2}h+\tanh^{2}h)] \nonumber\\ &\times
	\{\pi(1+\cosh4b+2\cosh2h)^{2}\,\text{sech}^{2}h-\text{sech}^{2}h-
	16 \,\bar{k} \nonumber \\ & \times\{4\text{sech}^{2}[\frac{1}{2}\log(\cosh^{2}2b \,\text{sech}^{2}h+
	\tanh^{2}h)]\times\tanh^{2}[\frac{1}{2}\log(\cosh^{2}2b \,\text{sech}^{2}h+\tanh^{2}h)]\}\nonumber \\ &\times
	(\cosh4b+\cosh2h-8\cosh^{4}b \,\text{sech}^{2}h\,\sinh {4b})\}
	\sinh4b, \label{eq:A.2} \tag{A.2}
\end{align}
\begin{align}
	\frac{\partial f_{2}}{\partial h}&=\frac{1}{16\pi(\cosh4b+\cosh2h)^{2}}\,
	\text{csch}[\frac{1}{2}\log(\cosh^{2}{2b}\,  \text{sech}^{2}h+\tanh^{2}h)]\, \nonumber \\ &\times
	\text{sech}[\frac{1}{2}\log(\cosh^{2}2b \, \text{sech}^{2}h+\tanh^{2}h)]
	\nonumber\\ &\times\{\pi(1+\cosh4b+2 \cosh2h)^{2}\,\text{sech}^{2}h-
	16\,\bar{k}\{4\text{sech}^{2}[\frac{1}{2}\log(\cosh^{2}2b \,\text{sech}^{2}h
	+\tanh^{2}h)] \nonumber\\ &\times\tanh^{2}[\frac{1}{2}\log(\cosh^{2}2b \, \text{sech}^{2}h
	+\tanh^{2}h)]\}(\cosh4b+\cosh2h-
	8\cosh^{4}b \,\text{sech}^{2}h\,\sinh{4b})\} \nonumber\\ & \times\sinh^{2}2b\tanh h. \label{eq:A.3} \tag{A.3}
\end{align}
where the parameter  $\bar{k}   $  defines  the  first type of the elliptic integral.


\begin{thebibliography}{00}
	\bibitem{Weinhold}
	F. Weinhold, Metric geometry of equilibrium thermodynamics, J. Chem. Phys. {\bfseries63}, 2479-2483 (1975). 
	\hyperref{https://doi.org/10.1063/1.431689}{}{}{	
		doi:10.1063/1.431689}
	\bibitem{Ruppeiner}
	G. Ruppeiner, Thermodynamics: A Riemannian geometric model, Phys. Rev. A {\bfseries20}, 1608-1613 (1979).  
	\hyperref{https://doi.org/10.1103/PhysRevA.20.1608}{}{}{	 doi:10.1103/PhysRevA.20.1608}
	\bibitem{geometry}
	G. Ruppeiner, Riemannian geometry in thermodynamic fluctuation theory, Rev. Mod. Phys. {\bfseries67}, 605 (1995). 
	\hyperref{https://doi.org/10.1103/RevModPhys.67.605
	}{}{}{  doi:10.1103/RevModPhys.67.605}
	\bibitem{Mirza}
	S. A. H. Mansoori and B. Mirza, Correspondence of phase transition points and singularities of thermodynamic geometry of black holes, Eur. Phys. J. C {\bfseries74}, 2681 (2014).
	\hyperref{https://doi.org/10.1140/epjc/s10052-013-2681-6
	}{}{}{	doi:10.1140/epjc/s10052-013-2681-6}
	\bibitem{Fazel}
	S. A. H. Mansoori, B. Mirza, and M. Fazel, Hessian matrix, specific heats, Nambu brackets, and thermodynamic geometry, JHEP {\bfseries2015}, 115 (2015).  
	\hyperref{https://doi.org/10.1007/JHEP04%282015%29115
	}{}{}{ doi:10.1007/JHEP04
	}
	\bibitem{Rupplist1}
	G. Ruppeiner, Thermodynamic curvature measures interactions, Am. J. Phys. {\bfseries78}, 1170 (2010). 
	\hyperref{https://doi.org/10.1119/1.3459936
	}{}{}{ doi:10.1119/1.3459936}
\bibitem{Janyszek}
H. Janyszek  and R. Mrugaa, Riemannian geometry and stability of ideal quantum gases, J. Phys. A:  Math. Gen. {\bfseries23}, 467 (1990).  
\hyperref{https://doi.org/10.1088/0305-4470/23/4/016
}{}{}{ doi:10.1088/0305-4470/23/4/016}
	\bibitem{anyon1}
B. Mirza and H. Mohammadzadeh, Ruppeiner geometry of Anyon gas, Phys. Rev. E {\bfseries78}, 021127 (2008). 
\hyperref{https://doi.org/10.1103/PhysRevE.78.021127 
}{}{}{ doi:10.1103/PhysRevE.78.021127}
\bibitem{Mohammadzadeh}
B.  Mirza and H. Mohammadzadeh, Nonperturbative thermodynamic geometry of anyon gas, Phys. Rev. E {\bfseries80}, 011132 (2009).
\hyperref{https://doi.org/10.1103/PhysRevE.80.011132
}{}{}{ doi:10.1103/PhysRevE.80.011132}

	\bibitem{s1}
	G. Ruppeiner, Application of Riemannian geometry to the thermodynamics of a simple
	fluctuating magnetic system, Phys. Rev. A {\bfseries24}, 488-492 (1981).  \hyperref{https://doi.org/10.1103/PhysRevA.24.488
	}{}{}{ doi:10.1103/PhysRevA.24.488}

\bibitem{kagome}
B. Mirza and Z. Talaei, Thermodynamic geometry of a Kagome Ising model in a magnetic field, Phys. Lett. A {\bfseries377}, 513 (2013).
\hyperref{https://doi.org/10.1016/j.physleta.2012.12.030}{}{}{	doi:10.1016/j.physleta.2012.12.030}
	\bibitem{s2}
	D. Brody and N. Rivier, Geometrical aspects of statistical mechanics, Phys. Rev. E {\bfseries51},1006-1011 (1995). \hyperref{https://doi.org/10.1103/PhysRevE.51.1006
	}{}{}{ doi:10.1103/PhysRevE.51.1006}
	\bibitem{s3}
	D. C. Brody and A. Ritz, Information geometry of finite Ising models, J. Geom. Phys. {\bfseries47}, 207-220 (2003). \hyperref{https://doi.org/10.1016/S0393-0440(02)00190-0
	}{}{}{  doi:10.1016/S0393-0440(02)00190-0}
	\bibitem{Nakamura}
	T. Nakamura,  Scalar Curvature of the Quantum Exponential Family for the Transverse-Field Ising Model and the Quantum Phase Transition, (2022). 
	\hyperref{https://doi.org/10.48550/arXiv.2212.12919
	}{}{}{ 
		doi:10.48550/arXiv.2212.12919}
	\bibitem{spheric}
	D. A. Johnston, W. Janke, and R. Kenna, Information geometry, one, two, three (and four), 
	Acta Phys. Pol. B {\bfseries34}, 4923-4937 (2003).  
	\hyperref{https://doi.org/10.48550/arXiv.cond-mat/0308316
	}{}{}{	doi:10.48550/arXiv.cond-mat/0308316}
	\bibitem{Geometrothermodynamics}
	S. A. H. Mansoori and B. Mirza, Geometrothermodynamics as a singular conformal thermodynamic geometry, Phys. Lett. B {\bfseries799}, 135040 (2019). 
	\hyperref{https://doi.org/10.1016/j.physletb.2019.135040}{}{}{	doi:10.1016/j.physletb.2019.135040}.
	\bibitem{blackhole2}
	B. Mirza, M. Zamaninasab, Ruppeiner geometry of RN black holes 
	flat or curved? 	JHEP {\bfseries1006}, 059 (2007).  
	\hyperref{https://doi.org/10.1088/1126-6708/2007/06/059}{}{}{	doi:10.1088/1126-6708/2007/06/059}
	\bibitem{kerr}
	G. Ruppeiner, Thermodynamic curvature and phase transitions in Kerr-Newman black
	holes, Phys. Rev. D {\bfseries78}, 024016-1-13 (2008). 
	\hyperref{https://doi.org/10.1103/PhysRevD.78.024016}{}{}{	doi:10.1103/PhysRevD.78.024016}
	\bibitem{Universal1}
	D. Kubiznak and R. B. Mann, P-V criticality of charged AdS black holes, J. High Energy Phys.  {\bfseries1207}, 033 (2012).  
	\hyperref{https://doi.org/10.1007/JHEP07(2012)033}{}{}{	doi:10.1007/JHEP07(2012)033} 
	\bibitem{hairyblackholes}
	B. Mirza and Z. Sherkatghanad, Phase transitions of hairy black holes in massive gravity and thermodynamic behavior of charged AdS black holes in an extended phase space, Phys. Rev. D {\bfseries90}, 084006 (2014). 
	\hyperref{https://doi.org/10.1103/PhysRevD.90.084006}{}{}{	doi:10.1103/PhysRevD.90.084006}
	\bibitem{Universal}
	S. A. H. Mansoori, M. Rafiee, and S.-W. Wei, Universal criticality of thermodynamic curvatures for charged AdS black holes, Phys. Rev. D {\bfseries102}, 124066 (2020). 
	\hyperref{https://doi.org/10.1103/PhysRevD.102.124066}{}{}{	 doi:10.1103/PhysRevD.102.124066}
	\bibitem{Mansoori}
	S. A. H. Mansoori, B. Mirza, and E. Sharifian, Extrinsic and intrinsic curvatures in thermodynamic geometry, Phys. Lett. B {\bfseries759}, 298-305 (2016). 
	\hyperref{https://doi.org/10.1016/j.physletb.2016.05.096}{}{}{	doi:10.1016/j.physletb.2016.05.096}
	
	\bibitem{Salamon}
	P. Salamon, et al. On the relation between entropy and energy versions of thermodynamic length, J. Chem. Phys.  {\bfseries80}, 436-437 (1984).
	\hyperref{https://doi.org/10.1063/1.446467}{}{}{	 doi:10.1063/1.446467}
	\bibitem{Wik}
	F. Wilczek,  Quantum mechanics of fractional-spin particles, Phys. Rev. Lett. {\bfseries49}, 957 (1982). \hyperref{https://doi.org/10.1103/PhysRevLett.49.957}{}{}{	doi:10.1103/PhysRevLett.49.957}
	\bibitem{an1}
	F. D. M. Haldane, Fractional statistics in arbitrary dimensions: a generalization of the Pauli principle, Phys. Rev. Lett. {\bfseries67}, 937 (1991). \hyperref{https://doi.org/10.1103/PhysRevLett.67.937}{}{}{	doi:10.1103/PhysRevLett.67.937}
	\bibitem{an2}
	J. M. Leinaas and J. Myrheim, On the theory of identical particles, Nuovo Cimento  Soc. Ital. Fis., {\bfseries37B}, 1 (1977). 
	\bibitem{Wu}
	Y.-S.  Wu, Statistical Distribution of Particles obeying Fractional Statistics, Phys. Rev. Lett.{\bfseries73}, 922 (1994).
	\hyperref{https://doi.org/10.48550/arXiv.cond-mat/9402008
	}{}{}{ doi:10.48550/arXiv.cond-mat/9402008}
	\bibitem{d1}
	C. N. Yang and C.P. Yang, Thermodynamics of a one?dimensional system of bosons with repulsive delta?function interaction, J. Math. Phys.{\bfseries10}, 1115 (1969). \hyperref{https://doi.org/10.1063/1.1664947
	}{}{}{ doi:10.1063/1.1664947}
	\bibitem{Can}
	G. S. Canright  and S. M. Girvin, Fractional Statistics: Quantum Possibilities
	in Two Dimensions, Science {\bfseries247}, 1197 (1990).
	\hyperref{https://doi.org/10.1126/science.247.4947.1197
	}{}{}{ doi:10.1126/science.247.4947.1197}
	\bibitem{27}
	M. Chaichian, R. Gozales Felipe, and C. Montonen, Statistics of q-oscillators, quons and relations to fractional statistics, J. Phys. A {\bfseries26}, 4017 (1993). 
	\hyperref{https://doi.org/10.1088/0305-4470/26/16/018
	}{}{}{ doi:10.1088/0305-4470/26/16/018}
	\bibitem{23}
	M. V. N. Murthy and R. Shankar, Thermodynamics of a one-dimensional ideal gas with fractional exclusion statistics, Phys Rev. Lett. {\bfseries73}, 3331 (1994). \hyperref{https://doi.org/10.1103/PhysRevLett.73.3331
	}{}{}{ doi:10.1103/PhysRevLett.73.3331}
	\bibitem{24}
	C. Nayak and F. Wilczek, Exclusion statistics: Low-temperature properties, fluctuations, duality, and applications, Phys. Rev. Lett. {\bfseries73}, 2740 (1994). 
	\hyperref{https://doi.org/10.1103/PhysRevLett.73.2740
	}{}{}{ doi:10.1103/PhysRevLett.73.2740}
	\bibitem{25}
	F. M. D. Pellegrino, G. G. N. Angilella, N. H. March and R. Pucci, Statistical correlations in an ideal gas of particles obeying fractional exclusion statistics, Phys. Rev. E
	{\bfseries76}, 061123 (2007). \hyperref{https://doi.org/10.1103/PhysRevE.76.061123
	}{}{}{ doi:10.1103/PhysRevE.76.061123}
	\bibitem{26}
	S. Vishveshwara, M. Stone and D. Sen, Correlators and fractional statistics in the quantum hall bulk, Phys. Rev. Lett. {\bfseries99}, 190401 (2007). 
	\hyperref{https://doi.org/10.1103/PhysRevLett.99.190401
	}{}{}{ doi:10.1103/PhysRevLett.99.190401}
	\bibitem{Huang1}
	W.-H. Huang, Boson-fermion transmutation and the statistics of anyons, Phys. Rev. E {\bfseries51}, 3729 (1995). 
	\hyperref{https://doi.org/abs/10.1103/PhysRevE.51.3729
	}{}{}{ doi:10.1103/PhysRevE.51.3729}
	\bibitem{Huang}
	W.-H. Huang,  Statistics of anyon gas and the factorizable property of thermodynamic quantities, Phy. Rev.  B {\bfseries53}, 15842 (1996).
	\hyperref{https://doi.org/10.1103/PhysRevB.53.15842
	}{}{}{ doi:10.1103/PhysRevB.53.15842}
	\bibitem{Huang3}
	W.-H. Huang, Comment on Quantum Statistical Mechanics of an Ideal Gas with Fractional Exclusion Statistics in Arbitrary Dimension, Phys. Rev. Lett. {\bfseries81}, 2392 (1998).
	\hyperref{https://doi.org/10.1103/PhysRevLett.81.2392
	}{}{}{ doi:10.1103/PhysRevLett.81.2392}.
	\bibitem{Baxter1966}
	G. Baxter, Weight Factors for the Two?Dimensional Ising Model, J. Math. Phys. {\bfseries6}:1015 (1965). 
	\hyperref{https://doi.org/abs/10.1063/1.1704362 	
	}{}{}{ doi:10.1063/1.1704362}
	\bibitem{Wuu}
	B. M. McCoy and T. T. Wu, Theory of Toeplitz determinants and the spin correlations of the two-dimensional Ising model. II, Phys. Rev. {\bfseries155 }(2): 438 (1967).
	\hyperref{https://doi.org/10.1103/PhysRev.155.438
	}{}{}{ doi:10.1103/PhysRev.155.438}
	\bibitem{fisher}
	M. E. Fisher, Lattice statistics in a magnetic field, I. A two-dimensional super-exchange antiferromagnet, Proc. of the Royal Society, A, {\bfseries254}, 66
	(1960).  
	\hyperref{https://doi.org/10.1098/rspa.1960.0005
	}{}{}{  doi:10.1098/rspa.1960.0005}
	\bibitem{Baxter}
	R. J. Baxter,  \textit{Exactly Solved Models in Statistical Mechanics}  (Academic: New York 1982 ).
	\bibitem{Azaria}
	P. Azaria and H. Giacomini, An exactly solvable two-dimensional Ising model with magnetic field, J. Phys. A {\bfseries21}, L935 (1988).
	\hyperref{https://doi.org/10.1088/0305-4470/21/19/003
	}{}{}{  doi:10.1088/0305-4470/21/19/003}
	\bibitem{Lin}
	K. Y. Lin, An exact result for the magnetisation of the Kagome lattice
	Ising model with magnetic field, J. Phys. A {\bfseries22}, 3435 (1989).  \hyperref{https://doi.org/10.1088/0305-4470/22/16/033
	}{}{}{ doi:10.1088/0305-4470/22/16/033}
	\bibitem{Lu}
	W. T. Lu, F. Y. Wu, Soluble kagome Ising model in a magnetic field, Phys. Rev. E {\bfseries71}, 046120 (2005). \hyperref{https://doi.org/10.1103/PhysRevE.71.046120
	}{}{}{  doi:10.1103/PhysRevE.71.046120}
	\bibitem{fhc2}
	R. M. F. Houtappel, Order-disorder in hexagonal lattices, Physica  {\bfseries16}, 425-455 (1950). \hyperref{https://doi.org/10.1016/0031-8914(50)90130-3
	}{}{}{  doi:10.1016/0031-8914(50)90130-3}
	\bibitem{fhc1}
	I. Syozi,  C. Domb and M. S. Green,  \textit{Phase Transitions and Critical Phenomena}  ((Academic Press, New York 1972).
	\bibitem{ebrahimi}
	M. E. Khuzani, B. Mirza, and M. T. Kachi, Thermodynamic geometry of pure Lovelock black holes, Int. J. Mod. Phys. D.  {\bfseries31},   2250097 (2022).  \hyperref{https://doi.org/10.1142/S0218271822500973}{}{}{	doi:10.1142/S0218271822500973}
	\bibitem{Kac}
	T. Berlin and M. Kac, The Spherical Model of a Ferromagnet, Phys. Rev. {\bfseries86} 821 (1952).  \hyperref{https://doi.org/10.1103/PhysRev.86.821
	}{}{}{ doi:10.1103/PhysRev.86.821}
	\bibitem{random}
	D. V. Boulatov and V.A. Kazakov, The ising model on a random planar lattice: The structure of the phase transition and the exact critical exponents , Phys. Lett. B  {\bfseries186} (1987). \hyperref{https://doi.org/10.1016/0370-2693(87)90312-1
	}{}{}{ doi:10.1016/0370-2693(87)90312-1}
	\bibitem{inform}
	W. Janke, D. A. Johnston, and Ranasinghe P.K.C. Malmini, The Information Geometry  of the Ising Model on Planar Random Graphs, Phys. Rev. E.  {\bfseries186} (2002).  \hyperref{https://doi.org/10.1103/PhysRevE.66.056119
	}{}{}{ doi:10.1103/PhysRevE.66.056119}
	\bibitem{spheric22}
	W. Janke, D. A. Johnston, and R. Kenna, Information geometry of the spherical model, Phys. Rev. E {\bfseries67}, 046106, (2003). \hyperref{https://doi.org/10.1103/PhysRevE.67.046106
	}{}{}{ doi:10.1103/PhysRevE.67.046106}


	\bibitem{yyans}
	H. Janyszek, Riemannian geometry and stability of thermodynamical
	equilibrium systems, J. Phys. A {\bfseries23}, 477 (1990). \hyperref{https://doi.org/10.1088/0305-4470/23/4/017
	}{}{}{ doi:10.1088/0305-4470/23/4/017}
\end{thebibliography}
\end{document}